\documentclass[ 
 twocolumn,%showpacs,preprintnumbers,
 amsmath,amssymb]{revtex4} % PRD--TT08
\usepackage[pdftex]{graphicx,color}           %------ for PDF-figure
\usepackage{ulem}                                        %------ Strikethrough

  \newcommand{\EE}{\hat{E}}
  \newcommand{\LL}{\hat{L}}
  \newcommand{\ee}{\hat{e}} 
   
 \newcommand{\aap}{Astron. Astrophys.}   % Astronomy and Astrophysics
 \newcommand{\mnras}{Mon. Not. R. Astron. Soc.} % Monthly Notices of the Royal Astronomical Society
 \newcommand{\pasj}{Publ. Astron. Soc. Jpn.}   % Publications of the Astronomical Society of the Japan
\begin{document}

\title{ Relativistic jet acceleration region in a black hole magnetosphere }
\author{ Masaaki Takahashi }%
\email{mtakahas@auecc.aichi-edu.ac.jp} 
\affiliation{% 
 Department of Physics and Astronomy, Aichi University of Education,  
 Kariya, Aichi 448-8542, Japan 
}
\author{ Motoki Kino }%
\email{motoki.kino@nao.ac.jp}
\affiliation{% 
 Kogakuin University of Technology \& Engineering, Academic Support Center, 2665-1 
 Nakano, Hachioji, Tokyo 192- 0015, Japan
}\affiliation{% 
 National Astronomical Observatory of Japan 2-21-1 Osawa, Mitaka, Tokyo, 181-8588, Japan
}
\author{ Hung-Yi Pu }%
\email{hypu@ntnu.edu.tw}
\affiliation{% 
National Taiwan Normal University (NTNU), No. 88, Sec. 4, Tingzhou Road, Taipei 116, Taiwan, R.O.C.
}
\affiliation{% 
Institute of Astronomy and Astrophysics, Academia Sinica, 11F of Astronomy-Mathematics Building, 
AS/NTU No. 1, Sec. 4, Roosevelt Rd, Taipei 10617, Taiwan, R.O.C. 
}
%\date{\today}             % Activate to display a given date or no date
%===============================================================

\begin{abstract}
 We discuss stationary and axisymmetric trans-magnetosonic outflows in the magnetosphere 
 of a rotating black hole (BH). Ejected plasma from the plasma source located near the BH  
 is accelerated far away to form a relativistic jet. In this study, the plasma acceleration efficiency 
 and conversion of fluid energy from electromagnetic energy are considered by employing the 
 trans-fast magnetosonic flow solution derived by Takahashi \& Tomimatsu (2008).    
 Considering the parameter dependence of magnetohydrodynamical flows,  we search for 
 the parameters of the trans-magnetosonic outflow solution to the recent M87 jet observations   
 and obtain the angular velocity values of the magnetic field line and angular momentum of 
 the outflow in the magnetized jet flow. Therefore, we estimate the locations of the outer light 
 surface, Alfv\'{e}n surface, and separation surface of the flow. We also discuss the 
 electromagnetic energy flux from the rotating BH (i.e., the Blandford--Znajek process), which 
 suggests that the energy extraction mechanism is effective for the M87 relativistic jet. 
\end{abstract}

% \pacs{\red 23.23.+x, 56.65.Dy}  %... dummy 

\maketitle %--PRD

\section{ Introduction } %================================

The system of a supermassive black hole (BH) with accreting matter is widely believed to be 
the central engine of the galactic nuclei, and relativistic jets are often observed. Recently, 
very-long baseline interferometry (VLBI) observations have revealed the configuration 
and velocity distribution of the M87 jet near the BH \cite{Mertens+16,Chatterjee+19}.  
The parsec-scale jet of the galaxy M87 is parabolic-like \cite{AN12,NA13}, 
and it becomes conical-like near the BH \cite{Hada+13}.  The radial profile of the jet 
velocity was observed by Park et al. \cite{Park+19} using the KVN and VERA Array (KaVA)
(KVN: Korean VLBI Network;  VERA: VLBI Exploration of Radio Astronomy).  
The rapid increase in velocity is observed approximately 100 to 1,000 times the BH's radius.  
This area is called the acceleration region and is considered to extend from slightly inside 
the outer light surface to the area several times the fast-magnetosonic surface.
Thus, detailed data for the region closer to the base of the jet were obtained.  
Exploring the base of the jet would advance our understanding of plasma near the BH.  

%%\smallskip  

\begin{figure}[t] % [htbp]	
\includegraphics[width=6cm,clip]{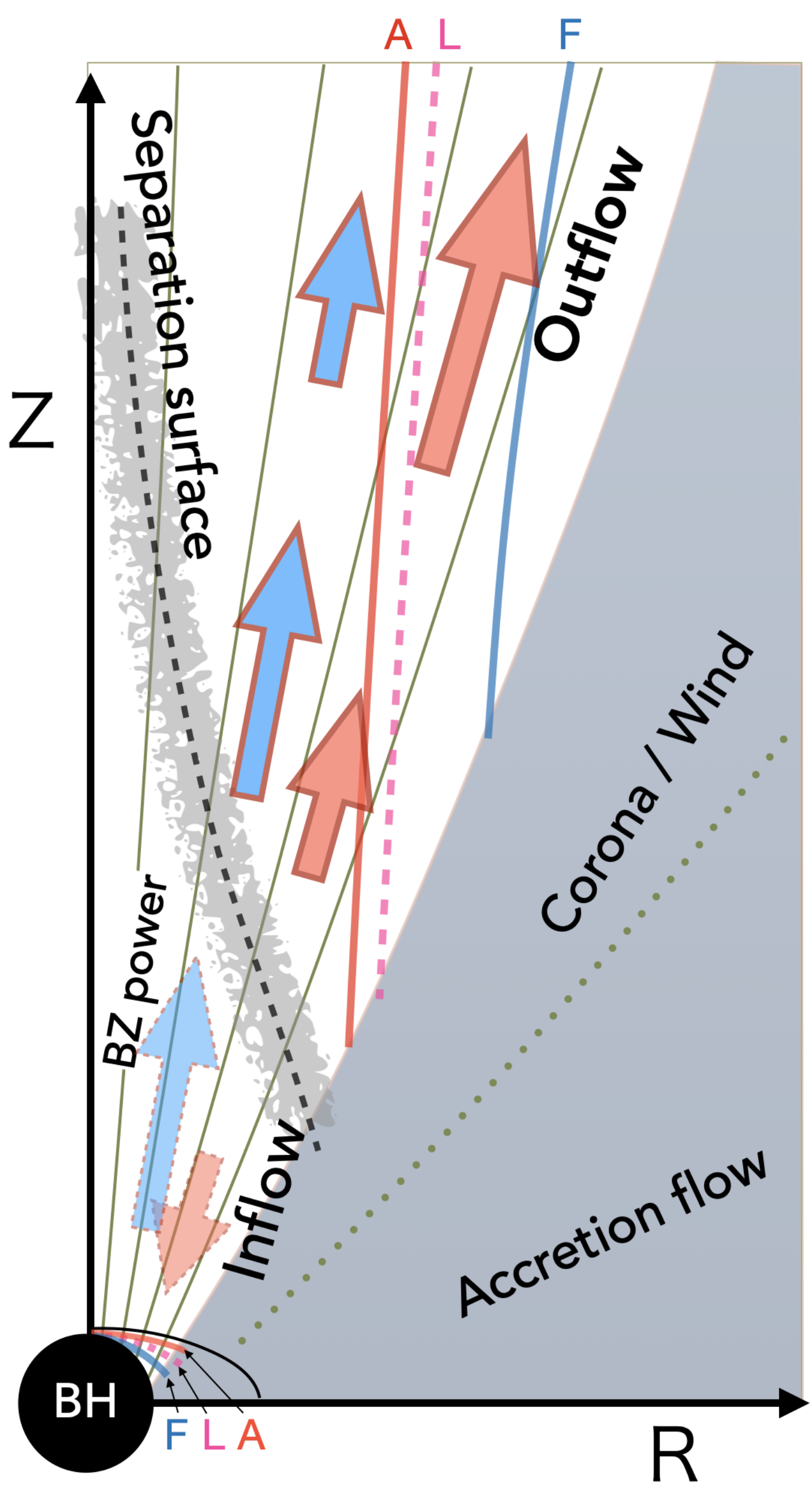}  
\caption{ 
    Schematic of MHD flows driven by rotating BH (BZ power). 
    The thin green curves are the global magnetic field lines in the magnetosphere.  
    The blue and red arrows indicate the direction of the Poynting and fluid component fluxes, 
    respectively. The ejected outflow and inflow from a plasma source, which may be near the 
    separation surface,  pass through the Alfv\'{e}n surface ({\sf A}), light surface ({\sf L}), and 
    fast-magnetosonic surface ({\sf F}) in order,  then accelerate toward a distant and the BH,  
    respectively.  The thin black curve around the BH is the boundary of the ergosphere. 
     }
 \label{fig:BH-jet} %%-- Figure_1  
\end{figure}

A magnetic field is generated by accretion plasma around the BH, which is dominant around 
the axis of rotation. Such a region is called the BH magnetosphere. 
 At the low- and mid-latitudes of this rotating system,  a gas torus is formed by accretion gas.  
The outflow is ejected from a plasma source located in the magnetosphere and accelerated 
by the Lorentz force toward a distant region. 
The BH magnetosphere extends along the flow from the BH to the region where the magnetic 
field and the fluid energies become comparable. We consider stationary and axisymmetric 
magnetohydrodynamic (MHD) outflows in the BH magnetosphere as relativistic jets.  
 Figure~\ref{fig:BH-jet} shows trans-fast magnetosonic outflow and inflow in a BH  
 magnetosphere. 
 The flow region (a funnel) is confined by a corona and/or disk wind, and cold MHD flow is 
 considered.  It is shown separately for the electromagnetic field component (Poynting flux: 
 blue arrows)  and fluid component (red arrows) of the MHD flow's energy flux.  
 The jet power derived from the rotational energy of the BH would be explained via the 
 Blandford--Znajek (BZ) process  \cite{BZ77,Znajek77}  (see also, \cite{LOP99,BK2000}).  
The rotation of the magnetosphere generates a strong centrifugal force on plasma, thereby 
creating a region that produces an outward plasma flow along a magnetic field line; moreover, 
there is a region of accretion flow because of strong gravity near the BH.  
Therefore, there is a plasma supply between the inflow and outflow regions,    
which is a watershed for inflow and outflow ejected from the plasma source with low velocity 
and is located between the inner and outer light surfaces (e.g., \cite{TNTT90,GL13}). 
Thus, the BH magnetosphere can be classified into inner and outer regions; i.e., 
the inner and outer BH magnetospheres.  
The region where gravity and centrifugal force act on the plasma along the streamline balance 
is called the ``separation surface,'' $r_{\rm sp}(\theta)$.    

%%\smallskip 

%------------------------------------------------------------------------------------------------------------
%------------------------------------------------------------------------------------------------------------//

For a stationary and axisymmetric ideal MHD flow solution in the BH magnetosphere, 
in addition to the parameters of BH spacetime and the function of the magnetic field shape 
$\Psi(r, \theta)$, the total energy $E(\Psi)$ and total angular momentum $L(\Psi)$ of the flow, 
the angular velocity of the magnetic field lines $\Omega_F(\Psi)$, and the particle number 
flux per magnetic flux tube $\eta(\Psi)$ should be specified. 
These parameters are conserved along the magnetic field line (i.e.,  the flow's streamline). 
By analyzing these parameter values for the outflow solution with a solution of magnetic 
field distribution, various jet morphologies and behaviors would be elucidated. 
To apply the flow solution to an observed relativistic jet, it is necessary to associate the 
physical quantity obtained from the observation with the model parameters as the 
field-aligned conserved quantities. 

%%\smallskip 

We examine the nature of the magnetized outflow and then the acceleration of the jet near 
its base region. Further, we compare the outflow solution with the M87 observed 
data in the framework of general relativity.  We expect this attempt to impose restrictions 
on unknown plasma sources and inflow onto the BH. 
Generally, solving the equation of motion for magnetized fluid involves analysis of critical 
conditions at the Alfv\'{e}n point as well as fast and slow magnetosonic points  
under a given magnetic field line. 
This analysis is complex and tedious (see, e.g., \cite{Takahashi02,TS98}).
However,  Takahashi \& Tomimatsu \cite{TT08} (hereinafter, TT08) analytically devised 
trans-magnetosonic flow solution without a complicated regularity condition analysis at the 
magnetosonic surfaces. In this study, we adopt the method of TT08 to analyze trans-fast 
magnetosonic flow solutions. This method facilitates the analysis of stationary 
trans-magnetosonic flow solutions; i.e., we can easily solve the MHD flows and discuss 
the distributions of plasma density, the flow velocity, and the ratio of the magnetization 
of the flow. Hence, we apply the flow solution to the M87 jet. 

%%\smallskip 

The asymptotic structures of the magnetic field and relativistic flow have been 
discussed (e.g, \cite{Li+92,BL94,AC93,Lyubarsky09}).  
Recently, Huang, Pan \& Yu \cite{HPY20} numerically discussed a self-consistent 
magnetosphere of a BH by solving the transfield equation (the relativistic 
Grad--Shafranov equation) \cite{Camenzind87,NTT91} (see also \cite{Beskin97}) 
with the poloidal equation and applied it to the M87 jet to explain the velocity distribution.   
Kino et al. \cite{Kino+21} discussed the flow acceleration of the M87 jet by employing an 
analytical model of Tomimatsu \& Takahashi \cite{TT03} (hereinafter, TT03), which solved 
the approximated transfield equation outside the outer light surface. 

%%\smallskip 

Pu \& Takahashi \cite{PT20} discussed trans-fast magnetosonic jets in a BH 
magnetosphere by using TT08, where the injection surface $r_{\rm inj}(\Psi)$ of flows was 
set to the location of the separation surface $r_{\rm sp}(\Psi)$, and the injected flow starts 
with a zero velocity, $u^r_{\rm inj}=0$. They performed parameter searches for 
$\Omega_F$, $E$, $\Psi$ and the BH spin dependencies. Notably, the angular 
momentum $L$ of outflow injected from the separation surface with zero velocity can be 
described by using other conserved quantities. 

%%\smallskip 

 In this study, we release such an initial setting for outflows by Pu \& Takahashi \cite{PT20} 
 that restricted the $L$ parameter, and search the dependence of the values of $E$, $L$  
 and $\Omega_F$ on the jet acceleration motivated by the M87 observations, where the 
 configuration of the magnetic field is considered to be of conical shape for some parameter 
 searches.  The parameter search of $L$ is to allow the flows to have a finite initial velocity 
 ($u^r_{\rm inj} > 0$) at the separation surface, $r_{\rm inj} = r_{\rm sp}$, or an initial zero 
 velocity ($u^r_{\rm inj} = 0$) ejected from the radius of $r_{\rm inj} \geq r_{\rm sp}$.   
 The specific location of $r_{\rm inj}(\theta)$ is still under debate in the BH magnetosphere 
 as the jet base.  It could be due to electron-positron pair-creation from the background 
 MeV photons \cite{Moscibrodzka11,Wong21,GL14a,GL14b} or modest acceleration 
 within a charge-starved region at the null surface or the separation surface 
 \cite{BIP92,HO98,HP16,LR11,BT15,Kisaka+20}).   
 Recently, the structure of a BH magnetosphere in consideration of the pair-creation near 
 the horizon was studied via general-relativistic particle-in-cell simulations 
 \cite{Parfrey+19, Crinquand+20}.  
 Thus,  the initial position and velocity ($r_{\rm inj}$, $u_{\rm inj}^r$), which are associated 
 with the angular momentum $L$, are expected to shift from $r_{\rm inj} = r_{\rm sp}$ with 
 $u_{\rm inj}^r=0$ to $r_{\rm inj} \neq r_{\rm sp}$ with $u_{\rm inj}^r \neq 0$. 

%%\smallskip 

The rest of this manuscript is organized as follows. 
In Section~II, we review the trans-magnetosonic flow solutions discussed by TT08. 
In Section~III,  we summarize the properties of cold trans-fast magnetosonic  flow solutions 
by analyzing  the field-aligned flow parameters. 
In Section~IV,  we discuss the outflow solution in detail for comparison with the jet observations. 
The application to the M87 jet is presented in Section~V, and we discuss it in Section~VI.  
Using the physical parameters of the outflow solution obtained by fitting with observed data,  
we estimate the values of $E$, $\Omega_F$, and $L$; i.e., the locations of the outer light 
surface, Alfv\'{e}n surface, and separation surface of the outflow.  Further,   
the observed M87 jet power can be explained by the BZ power. 
The stationary model of magnetically driven jets in a BH magnetosphere 
would explain the observed properties of the M87 jet and similar low luminosity AGN jets.  
Finally, we give concluding remarks in Section~VII.

\section{ Basic equations for Trans-fast magnetosonic flow } %==============

 We assume stationary and axisymmetric ideal MHD flows in a BH magnetosphere 
 in Kerr geometry.  The background metric is written by the Boyer--Lindquist coordinates 
 with $c=G=1$,
 \begin{eqnarray}
  ds^2 &=& \left( 1-\frac{2mr}{\Sigma} \right) dt^2 
               + \frac{4amr\sin^2\theta}{\Sigma} dt d\phi             \nonumber  \\
           & & - \frac{{\mathcal A}\sin^2\theta}{\Sigma} d\phi^2
                - \frac{\Sigma}{\Delta} dr^2 - \Sigma d\theta^2  \ ,   
 \end{eqnarray}
 where $m$ and $a$ denote the mass and angular momentum per unit mass of the 
 BH, respectively, and 
 $\Delta \equiv r^2 -2mr +a^2 $, $\Sigma \equiv r^2 +a^2\cos^2\theta$, 
 ${\mathcal A} \equiv (r^2+a^2)^2 - \Delta a^2 \sin^2\theta$.  
 The particle number conservation is $(nu^\mu)_{;\mu} = 0$, where $n$ is the number 
 density of the plasma, and $u^\mu$ is the fluid 4-velocity.  The ideal MHD condition is 
 $u^\nu F_{\mu\nu} = 0$, where $F_{\mu\nu}$ is the electromagnetic tensor, and it is 
 written as 
 $F_{\mu\nu}   %% = A_{\nu;\mu} - A_{\mu;\nu} 
    = A_{\nu,\mu} - A_{\mu,\nu}$ using the vector potential $A_\mu$.   
 The relativistic polytropic relation is $P = K\rho_0^\Gamma$, where $\Gamma$ is the 
 adiabatic index, $K$ is related to the entropy, $\rho_0 = nm_{\rm part}$ is the rest mass 
 density, and $m_{\rm part}$ is the mass of plasma particles. 
 The equation of motion is ${T^{\mu\nu}}_{;\nu} = 0$. The energy--momentum tensor is given 
 by $T^{\mu\nu} = T^{\mu\nu}_{\rm fluid} + T^{\mu\nu}_{\rm EM}$, where the fluid part  is 
 $T^{\mu\nu}_{\rm fluid} = n\mu u^\mu u^\nu - Pg^{\mu\nu}$, the electromagnetic part is 
 $T^{\mu\nu}_{\rm EM} = (1/4\pi)[ {F^\mu}_{\lambda}F^{\lambda\nu} +(1/4)g^{\mu\nu}F^2 ]$,  
 $\mu \equiv (\rho+P)/n$ is the relativistic enthalpy, $\rho$ is the total energy density, and 
 $F^2 \equiv F^{\mu\nu}F_{\mu\nu}$.  

 %%\smallskip 

\subsection{ Relativistic Bernoulli Equation for MHD Flows } \label{sec:R-Bernoulli} 
 
 We define the magnetic and electric fields as follows: 
 $B_\mu \equiv \! ~^{\ast}F_{\mu\nu}k^{\nu}$ and $E_\mu \equiv
   F_{\mu\nu}k^{\nu}$, where $^{\ast}F_{\mu\nu}\equiv (1/2)\sqrt{-g}
   \varepsilon_{\mu\nu\sigma\lambda}F^{\sigma\lambda}$, 
 and $k^\nu=(1,0,0,0)$ is the timelike Killing vector.  The poloidal component $B_p$ of 
 the magnetic field seen by a lab-frame observer is given by     
 \begin{eqnarray}
   B_p^2 &\equiv&   %% - B^A B_A / G_t^2               \nonumber  \\ 
               - (B^r B_r + B^\theta B_\theta) / G_t^2       \nonumber  \\ 
         &=&      - \left[ g^{rr}(\partial_r \Psi)^2  
               + g^{\theta\theta}(\partial_\theta \Psi)^2 \right]/{\rho_w^2}    \ , 
 \end{eqnarray}
 where  %%--[\refA]--  the index $A$ indicates the poloidal components ($A = r,\theta$) and
 $G_t \equiv g_{tt} + g_{t\phi}\Omega_F$ and 
 $\rho_w^2 \equiv g_{t\phi}^2 - g_{tt}g_{\phi\phi} = \Delta \sin^2 \theta$. 
 The function $\Psi(r,\theta)$ is the magnetic stream function (the $\phi$ component 
 of the vector potential, $A_\phi(r,\theta)$).  
 The ideal MHD flow stream along the magnetic field lines, $\Psi(r,\theta) =$ constant lines, 
 and have five field-aligned flow parameters;  the particle number flux per magnetic flux 
 $\eta(\Psi) = n u_p/B_p$,  total energy 
 $E(\Psi) = \mu u_t - \Omega_FB_\phi/(4\pi\eta)$,  total angular momentum 
 $L(\Psi) = - \mu u_\phi -B_\phi/(4\pi\eta)$, entropy,  %% which is related to $K(\Psi)$,  
  and angular  velocity of magnetic field lines  $\Omega_F(\Psi) = - F_{tA}/F_{\phi A}$ 
 \cite{BO78,Camenzind86a}.    
 Thus, we can treat one-dimensional MHD flows along a $\Psi(r,\theta) =$ constant line. 
 The physical variables of flows are denoted as a function of $r$ and $\Psi$ with 
 field-aligned flow parameters.  

%%\smallskip 

 We define the poloidal velocity $u_p$ by  
 $u_p^2 \equiv - (u^r u_r + u^\theta u_\theta)$.  %%--- $u_p^2\equiv -u_Au^A$.  
 The relativistic  Alfv\'{e}n Mach number $M$ is defined by  
\begin{equation}
      M^2\equiv \frac{4\pi\mu n u_p^2}{B_p^2} 
              = \frac{\hat{\mu} u_p}{{\cal B}_p} \ ,   \label{eq:mach_def}
\end{equation}
 where the term ${\cal B}_p \equiv B_p/(4\pi\mu_{c}\eta) $ is introduced to 
 nondimensionalize and $\mu_{c}$ is the enthalpy for a cold flow; i.e.,   
 $\mu_{c}=m_{\rm part}$.  The nondimensional toroidal component of the 
 magnetic field ${\cal B}_\phi$ is defined as 
\begin{equation}
  {\cal B}_\phi \equiv 
      \left( \frac{1}{\rho_w} \right) \frac{B_\phi}{4\pi\mu_{c}\eta}  
      = \frac{G_\phi{\hat E} + G_t{\hat L}}{\rho_w(M^2-\alpha)} \ ,              \label{eq:Bf}
\end{equation}
 where  ${\hat E}\equiv E/m_{\rm part}$, ${\hat L}\equiv L/m_{\rm part}$, 
 $G_\phi \equiv g_{t\phi} + g_{\phi\phi}\Omega_F$, and 
 $\alpha \equiv G_{t} + G_{\phi}\Omega_F$. The toroidal magnetic field 
 $B_\phi = (\Delta/\Sigma)F_{\theta r}$  can be expressed in terms of the field-aligned 
 flow parameters and Alfv\'en Mach number. The point of $M^2 = \alpha$ with 
 $G_\phi \hat{E} + G_t \hat{L} = 0$ on the flow solution is called the ``Alfv\'{e}n point'' 
 (labeled as ``A''); i.e., the point $(r_{\rm A}, M^2_{\rm A}; \Psi)$ in the 
 $r$--$M^2$ plane.  At the Alfv\'{e}n point, $B_{\phi{\rm A}}$ has a nonzero finite value.  
 The radius $r_{\rm A}$ is called the ``Alfv\'{e}n radius.''  
 Notably, we may identify the radius given by $G_\phi\hat{E} + G_t\hat{L} = 0$ without 
 $M^2=\alpha$. Such a radius is called an ``anchor radius'' for the magnetic field line 
 considered (TT08), where $B_{\phi} = 0$.  Although the magnetic field line in an 
 axisymmetric magnetic surface (i.e., $\Psi =$ constant surface) has a spiral shape 
 in the toroidal direction,  the direction of winding is reversed at the anchor radius if 
 such a radius appears on the magnetic surface. Both the Alfv\'{e}n and anchor radii 
 can be located between the inner and outer light  surfaces (i.e., the 
 $\alpha(r,\theta) > 0$ region), where the two light surfaces are given by the locations 
 $\alpha(r, \Psi; \Omega_F, a) = 0$. 

%% \smallskip 

 The relativistic Bernoulli equation for MHD flows, which determines the poloidal velocity 
 (or the Alfv\'{e}n Mach number) along a magnetic field line, can be expressed as follows:  
 \cite{Camenzind86b,TNTT90,TT08}
 \begin{equation}
     \ee^2 = \hat{\mu}^2 \alpha + M^4 (\alpha {\mathcal B}_p^2 + {\mathcal B}_\phi^2) \ , 
                                                    \label{eq:pol_eq}
 \end{equation}
 where $\ee\equiv \EE -\Omega_F \LL$.
 The relativistic enthalpy $\hat{\mu}$ is expressed in terms of $u_p$ and $B_p$
\begin{equation}
      \hat{\mu} \equiv \frac{\mu}{m_{\rm part}}  
        = 1 + \left( \frac{\mu_{\rm inj}}{m_{\rm part}} \right) \left(\frac{B_p}{u_p} \right)^{\Gamma-1}  
        = 1 + \mu_{\rm hot}\left( \frac{{\cal B}_p}{u_p} \right)^{\Gamma-1}  \ ,  \label{eq:enthalpy_eq}
\end{equation}
 where 
 $ \mu_{\rm hot} \equiv (\mu_{\rm inj}/m_{\rm part})(4\pi\mu_{c}\eta)^{\Gamma-1}$, 
 and the term $\mu_{\rm inj}$ is evaluated at the plasma injection point by  
\begin{equation}
       \mu_{\rm inj}  \equiv \frac{\Gamma K}{\Gamma-1}(m_{\rm part}\eta)^{\Gamma-1}   
        =  \frac{\Gamma}{\Gamma-1} \frac{P_{\rm inj}}{n_{\rm inj}m_{\rm part}}  
            \left(\frac{u_p^{\rm inj}}{B_p^{\rm inj}} \right)^{\Gamma-1} \ . 
\end{equation}

%%\smallskip

 Now,  to solve the relativistic Bernoulli equation (\ref{eq:pol_eq}),  
 we introduce the function $\beta(r; \Psi)$ as follows: \cite{TT08} 
\begin{equation}
     \beta \equiv \frac{ {\cal B}_\phi }{ {\cal B}_p }  \ ,  
\end{equation}
i.e., the pitch angle of a magnetic field line on a magnetic flux surface. Moreover,   
we can introduce the poloidal electric-to-toroidal magnetic field amplitude ratio 
$\xi(r,\theta) ~ \equiv \bar{E}_p / | \bar{B}_T | $,  
seen by a zero angular momentum observer (ZAMO),   
where $\bar{E}_p^2 \equiv - (\bar{E}_r \bar{E}^r + \bar{E}_\theta \bar{E}^\theta)$ 
and $\bar{B}_T^2 \equiv - \bar{B}_\phi \bar{B}^\phi$. The electric and magnetic fields 
in ZAMO are $\bar{E}_\alpha \equiv F_{\alpha\beta} h^\beta$ and  
$\bar{B}_\alpha \equiv (1/2)\eta_{\alpha\beta\gamma\delta} h^\beta F^{\gamma\delta}$ 
with $h^\beta = (\Sigma\Delta/{\cal A})^{-1/2}(1,0,0,\omega)$.  
The function $\xi(r,\theta)$ is related to the above function $\beta(r,\theta)$, as 
$ \xi^2 = - g_{\phi\phi} ( \Omega_F - \omega )^2 / \beta^2 $,  
where $\omega \equiv -g_{t\phi} / g_{\phi\phi}$.  

%%\smallskip

 By assuming the functions of the magnetic field line $\Psi=\Psi(r, \theta)$ 
 and $\beta=\beta(r,\theta)$ [or $\xi=\xi(r,\theta)$],  we can specify the cross-section 
 of the magnetic flux tube of the MHD flow.   
 Hence, using Eq.~(\ref{eq:Bf}),  Eq. ~(\ref{eq:pol_eq}) can be expressed 
 in terms of $M^2$ and $\beta$ with the conserved quantities.   
 Although the ratio $\beta$ should be determined by solving the transfield equation 
 with the poloidal equation self-consistently,  in this study, we only consider a magnetic 
 flux tube with a certain functional form of  $\beta(r,\theta)$.  
 Notably, the function $\beta(r; \Psi)$ must satisfy certain restrictions at several 
 characteristic locations in a BH magnetosphere (TT08).  

%%\smallskip

\subsection{ Cold MHD Flow Solutions }

The Alfv\'{e}n Mach number and poloidal velocity along the flow are obtained by solving 
Eq.~(\ref{eq:pol_eq}) with Eq.~(\ref{eq:Bf}), which is a higher-order equation for $M^2$ 
(or $u_p$).  However, for a cold MHD flow ($P = 0$), we obtain the following quadratic 
equation for $M^2$: \cite{TT08}  
\begin{equation}
     A M^4 - 2B M^2 + C = 0   ~ ,                  \label{eq:poloidal-M2-cold}
\end{equation} 
where 
\begin{eqnarray}
   %%   A &=& {\brown \left( z - \frac{1}{\beta^2} \right) \frac{1}{\rho_w^2}
   %%                            \left( G_\phi \hat{E} + G_t \hat{L} \right)^2 ~ , } \\ 
    A &=& - (k+1) - \frac{1}{\beta^2 \rho_w^2} \left( G_\phi \hat{E} + G_t \hat{L} \right)^2 ~ , \\ 
    B &=& \hat{e}^2 - \alpha ~ , \\ 
    C &=& \alpha \left( \hat{e}^2 -\alpha \right) ~,     
\end{eqnarray}
with   %%  $ z \equiv - (k+1) \rho_w^2 / (G_\phi \hat{E} + G_t \hat{L})^2$ and 
$ k \equiv (g_{\phi\phi}\hat{E}^2 + 2g_{t\phi}\hat{E}\hat{L} + g_{tt}\hat{L}^2)/\rho_w^2$. 
The Alfv\'{e}n Mach number of the flow is obtained using the following:  
\begin{equation}
   M_{(\pm)}^2 = \frac{ B \pm \sqrt{ B^2 - AC } }{A} ~,    \label{eq:poloidal-M2-pm}
\end{equation}
where $M_{(+)}^2 \, ( > \alpha)$ denotes the super-Alfv\'{e}nic flow, and 
$M_{(-)}^2 \, ( < \alpha)$ denotes the sub-Alfv\'{e}nic flow. At the Alfv\'{e}n point, 
we see the condition $\tilde{L}\Omega_F = Y_{\rm A}$ with  the definitions 
$\tilde{L} \equiv \hat{L}/\hat{E}$ and $ Y(r,\theta) \equiv - G_\phi \Omega_F / G_t$; 
therefore, we have $(B^2-AC)_{\rm A}=0$, and we confirm that the Alfv\'{e}n Mach number 
becomes $M_{\rm A}^2 \equiv M^2_{(+){\rm A}}  =M^2_{(-){\rm A}} = \alpha_{\rm A}$.  
At the light surface, we have $M^2_{(-){\rm L}} = 0$ and 
$M^2_{(+){\rm L}}  = 2 \hat{e}^2 / [\, - (k_{\rm L}+1) - \hat{e}^2/\beta^2_{\rm L} \,]  $. 
  
%%\smallskip 
  
For the cold MHD flow, we define two characteristic Alfv\'{e}n Mach numbers related to 
the Alfv\'{e}n and fast-magnetosonic wave speeds,  
$M^2_{\rm AW}(r, \theta) \equiv \alpha(r,\theta)$, and 
$M^2_{\rm FW}(r, \theta) \equiv \alpha(r,\theta) + \beta^2(r,\theta)$, respectively  
\cite{TNTT90,TT08}.  
At the light surface, we obtain $M_{\rm AW}^2(r_{\rm L}; \Psi) = 0$ and  
$M_{\rm FW}^2(r_{\rm L}; \Psi) = \beta^2$, and at the event horizon 
$r_{\rm H} = m + \sqrt{m^2 - a^2}$,  we have 
$M_{\rm AW}^2(r_{\rm H}; \Psi) = g_{\phi\phi} (\Omega_F - \omega_{\rm H})^2 < 0 $ 
and $M_{\rm FW}^2(r_{\rm H}; \Psi) = 0$, where $\omega_{\rm H} = a/(2mr_{\rm H})$ is 
the angular velocity of the BH. Hence, the inflow into the BH must be a trans-fast 
magnetosonic flow between the Alfv\'{e}n surface and the event horizon. The location of 
$(\alpha)'_{\rm sp} = 0$, where the prime is a derivative along a streamline 
$(~)' \equiv \partial_r + (B^\theta/B^r) \partial_\theta$,  makes the separation surface 
$r = r_{\rm sp}(\theta)$. 

%% \smallskip 

 For a trans-Alfv\'enic outflow, we should select $ M^2 = M^2_{-}  \,  ( < M_{\rm AW}^2) $ 
 in the sub-Alfv\'{e}nic region of $\tilde{L}\Omega_F > Y$, where is the region from the 
 injection point to the outer Alfv\'en point,  whereas $ M^2 = M^2_{+}  \,  ( > M_{\rm AW}^2) $ 
 is in the super-Alfv\'{e}nic region of $\tilde{L}\Omega_F < Y$, where is the region from the 
 outer Alfv\'en point to a distant region. The solution of $ M^2 = M^2_{+}$ in the 
 super-Alfv\'{e}nic region becomes trans-fast magnetosonic.  
 Notably, at the fast-magnetosonic point (labeled as ``F''), the solution of the above quadratic 
 equation does not diverge provided $\beta(r;\Psi)$ is a smooth function such that we can obtain 
 a trans-fast  magnetosonic solution without the usual critical analysis at the magnetosonic point.   
 However, if there is a location $A = 0$ along a distant region, the Mach number $M^2$ 
 begins to diverge there, where the particle number density and the magnetization parameter  
 become zero. To obtain a physical flow solution ejected from the plasma source, we require 
 the condition $A(r) > 0$ in all areas of the super-Alfv\'{e}nic flow.   

%%\smallskip 

Using the definition of the Alfv\'{e}n Mach number,  the poloidal velocity of the cold flow 
$u_p$ is given by 
\begin{equation}
     u_p^2 = {\cal B}^2_p \, M^4 = \frac{1}{ \beta^2 }\, 
     \frac{ (G_\phi \hat{E} + G_t \hat{L} )^2 }{\rho_w^2 (M^2 - \alpha)^2 }\, M^4  ~ ,   
    \label{eq:pol_velocity-1}
\end{equation}
or using the poloidal Eq.~(\ref{eq:pol_eq}), we obtain  
\begin{equation}
    u_p^2 = \frac{ \hat{e}^2 - \alpha }{ \alpha + \beta^2 } ~.    \label{eq:pol_velocity}
\end{equation}
At the light surface, the poloidal velocity is $u_p = \hat{e}/{\beta}$, where we should select  
$M^2(r_{\rm L}) = M_{(+)}^2(r_{\rm L})$.  Notably, $M_{(-)}^2(r_{\rm L}) = 0$ with 
${\cal B}_p(r_{\rm L}) = 0$; i.e., the solution $M^2(r) = M_{(-)}^2(r)$ breaks at the light surface.    
At the Alfv\'{e}n and fast-magnetosonic surfaces, we have 
$u_p^2(r_{\rm A}) = u_{\rm AW}^2(r_{\rm A})$ and 
$u_p^2(r_{\rm F}) = u_{\rm FW}^2(r_{\rm F})$, respectively, where 
the 4-velocity of the Alfv\'{e}n and  fast-magnetosonic waves are defined as 
$ u_{\rm AW}^2 \equiv {\cal B}_p^2 M_{\rm AW}^4 = ({\cal B}_\phi^2 / \beta^2) \, \alpha^2 $, and 
$ u_{\rm FW}^2 \equiv {\cal B}_p^2 M_{\rm FW}^4 = ({\cal B}_\phi^2 / \beta^2) \, (\alpha + \beta^2)^2 $.  

%%\smallskip 

From defining the Alfv\'{e}n Much number, the distribution of the particle number density 
is obtained as follows:   
\begin{equation}
      n = \frac{ 4\pi \mu_c \eta^2 }{ M^2 } ~ . 
\end{equation}
The magnetization parameter $\sigma$, which is defined as the ratio of the Poynting flux to the 
total mass-energy flux seen by a ZAMO,  is expressed as follows: \cite{TRFT02,TGFRT06} 
\begin{equation}
   \sigma(r, \theta) = \frac{ B_\phi G_\phi }{ 4\pi \mu\eta u^t \rho_w^2 } 
   = - \frac{ \hat{e} - \alpha \hat{h} }{ \hat{e} - M^2 \hat{h} }  ~, 
\end{equation}
where $ \hat{h} \equiv g^{tt} ( \hat{E} -  \hat{L} \omega ) $. 
Thus, the energy conversion between the magnetic and fluid parts of the MHD flow's energy 
$E$ along the streamline is  discussed depending on plasma acceleration. 

%%\smallskip 

%------------------------------------------------------------------------------------------------------------
\begin{table*}[htb]   
  \caption{ 
  Classification of the trans-Alfv\'enic  flows and the relation between the flow's total energy 
  $E$ and angular momentum $L$. The condition $A(r;\Psi)>0$ in the super-Alfv\'{e}nic region 
  is considered.       
  }
  \begin{tabular}{|l||c|c|c||ccl| } \hline 
        &  $ \Omega_F^{\rm min} < \Omega_F \leq 0 $
        &  ~~$ 0 < \Omega_F < \omega_{\rm H} $~~
        &  $ \omega_{\rm H} \leq \Omega_F < \Omega_F^{\rm max}  $     &     &     &  \\ \hline \hline
    type I    &  $\cdots$ &  $\cdots$  &  ~$ (\tilde{L}\Omega_F)_{\rm min} < \tilde{L}\Omega_F \leq 1$  ~ 
                 &  ~$E>0$, & $L>0$~ & inflow or outflow ~  \\ \hline
    type IIa &  $\cdots$  &  $0 < \tilde{L}\Omega_F \leq 1$           &  $\cdots$
                 &  ~$E > 0$, & $L > 0$~   &   outflow         \\ \hline
    type IIb &   $\cdots$ & ~~~~ $\tilde{L}\Omega_F \leq 0$         &  $\cdots$
                 &  $~E \geq 0$, & $L \leq 0$~  &  inflow  \\ \hline
    type IIc &  $\cdots$  &  $1 \leq \tilde{L}\Omega_F$  ~~~~       &  $\cdots$ 
                 &  ~$E \leq 0$, & $L < 0$~  & inflow       \\ \hline
    type III  &   ~$ (\tilde{L}\Omega_F)_{\rm min} < \tilde{L}\Omega_F \leq 1$~   &  $\cdots$ & $\cdots$
                 &  ~$E > 0$, & $L < 0$~   &  inflow or outflow \\ \hline
  \end{tabular}
  \label{tab:alfven}
\end{table*}
%------------------------------------------------------------------------------------------------------------

%% medskip 

\section{ Properties of Trans-fast magnetosonic flows } %===========================
\label{sec:trans-fast jet}

Before considering the trans-fast magnetosonic outflow solution, the classification and constraints 
of the entire trans-fast magnetosonic inflow and outflow solutions in the BH magnetosphere is 
summarized.  
   
%% \smallskip

 For stationary and axisymmetric ideal MHD flows in the cold limit,  there are four conserved 
 quantities along the flow. The flow properties are characterized by these parameters with 
 some constraints.  In this section, we investigate the characteristics of the trans-fast 
 magnetosonic flow solution under the constraints. 
  
%% \smallskip 

\subsection{ Classification of the trans-Alfv\'enic  Flows }  
   
The magnetic field lines in the magnetosphere around the Kerr BH are dragged toward 
the rotation of the BH by the dragging effect of spacetime. In particular,  
the toroidal component of the magnetic field of $B_\phi \propto (\Omega_F-\omega)$ and 
the angular momentum of the flows $L \propto (\Omega_F - \omega_{\rm A})$ are strongly 
influenced by the drag of spacetime.  
The value of the angular momentum of the flow can be specified by the location of the 
Alfv\'{e}n point under given $E$ and $\Omega_F$. Trans-Alfv\'{e}nic MHD flows in a BH 
magnetosphere can be classified by $\tilde{L}\Omega_F$, which gives the constraint 
on the field-aligned flow parameters at the Alfv\'{e}n point by the relation 
$\tilde{L}\Omega_F = Y_{\rm A} $.         %%  \cite{TNTT90}.  

%% \smallskip 

The conditions at the Alfv\'{e}n point can be typed by the spin of a BH $a$ 
and the angular velocity of the magnetic field lines $\Omega_F$; i.e., 
  type~I ($ \omega_{\rm H}  \leq \Omega_F < \Omega_F^{\rm max} $), 
  type~II ($ 0  < \Omega_F < \Omega_{\rm H} $), and 
  type~III ($ \Omega_F^{\rm min} < \Omega_F \leq 0$), 
where $\Omega_F^{\rm min/max}$ is the minimum/maximum values of $\Omega_F$ for 
the existence of the inner and outer light surfaces. For types~I and III, 
where $(\tilde{L}\Omega_F)_{\rm min} < \tilde{L}\Omega_F \leq 1 $ with 
$E > 0$ and  $L > 0$ for type~I and $L < 0$ for type~III,  
both the inner and outer Alfv\'{e}n radii appear in the MHD flow solution.  
For type~II, however, there is one Alfv\'{e}n radius (the inner or outer one) in the solution, 
and it can be further classified into the following three cases: 
(a) type~IIa: $ 0 < \tilde{L}\Omega_F \leq 1$ with $L > 0$ and $E > 0$, 
(b) type~IIb: $ \tilde{L}\Omega_F \leq 0$ with $L \leq 0$ and $E \geq 0$ and  
(c) type~IIc: $ 1 \leq \tilde{L}\Omega_F $ with $L < 0$ and $E \leq 0$ 
(see also, TABLE~\ref{tab:alfven}).  

%% \smallskip 

\subsection{ Constraints on Trans-fast Magnetosonic Outflows }  

For the relativistic outflow such as the M87 jet, we will handle the flow through the Alfv\'{e}n 
surface located outside the separation surface, whch is called the outer Alfv\'{e}n surface.  
Such a flow has a positive angular momentum ($L>0$). 
Hence, we will consider the types~I and type~IIa cases. 
The steady magnetosonic flow in the BH magnetosphere has several essential surfaces  
that characterize its behavior. They include the separation surface, Alfv\'{e}n surface, 
light surface, and fast-magnetosonic surface. The outflow ejected from the plasma source 
passes through these surfaces before reaching the far distant region. 

%% \smallskip 

The locations of the Alfv\'{e}n surface are related to the field-aligned parameters as 
$\tilde{L} \Omega_F = Y_{\rm A}(\Psi)$,  and as a condition that the trans-Alfv\'{e}nic 
outflow reaches to a far distant region, $A (r,\theta) > 0$ should be required along the 
flow ($ r_{\rm A} < r <\infty$), as mentioned in the previous section.  
That is, at the outer Alfv\'{e}n surface $r = r_{\rm A}$, we find the restriction 
\begin{equation}
    A (r_{\rm A}; \Psi) = - k_{\rm A} -1 > 0 ~.
\end{equation} 
This imposes a limit on the  field-aligned flow parameters.  Thus, we have the condition 
\begin{equation}
   M^2_{\rm A} = \alpha_{\rm A}  > \frac{ G_{t{\rm A}}^2 }{ \hat{E}^2 } ~, 
\end{equation}
and we obtain the minimum energy for the trans-Alfv\'{e}nic outflows,   
\begin{equation}
   \hat{E}^2 > \frac{ G_{t{\rm A}}^2 }{ \alpha_{\rm A} } 
   = \frac{ G_{t{\rm A}} }{ 1 - \tilde{L}\Omega_F }  \equiv \hat{E}^2_{\rm min} ~.   \label{eq:E_min}
\end{equation}
Although the function $A(r;\Psi)$ includes the function $\xi^2$, the condition for the minimum 
energy (\ref{eq:E_min}) does not depend on the detail expression of $\xi^2(r,\theta)$; i.e., 
the minimum energy for jets does not limit the shape of the magnetic field as 
$\xi^2_{\rm A}(r,\theta)$. Notably, depending on the functional form of $\xi^2$, there may 
be a situation where $A = 0$ with $M^2\to\infty$ in the radius to infinity.   
For a trans-Alv\'{e}nic outflow solution,  if $A = 0$ before reaching the distant region,        
it becomes unphysical, unless MHD shock \cite{TRFT02,TGFRT06,TakahashiMR10}  
or some type of MHD instability (e.g., \cite{TMT01}) occurs in the flow.  
As discussed in TT03, the condition $\xi^2(r,\theta) < 1 - (1/\hat{E}^2) \equiv \xi_{\rm cr}^2$  
is required to reach a far distant region. 
In the BH magnetosphere, the condition that the radius of $A = 0$ does not appear in 
the trans-fast magnetosonic solution requires restrictions on $\tilde{L}\Omega_F$ and $\xi^2$. 
At the location of the $A = 0$ surface,  we have the following functions     %% [see, TT08] 
 \begin{eqnarray} %% {equation}
   ( \tilde{L}\Omega_F )^\pm(r; \Psi) &=& \frac{ Y }{ 1 + Y + X } 
     \biggl\{ 1 + X \pm \biggl[ 1 + (1-Y)X     \nonumber  \\
    & &  - (1+Y+X)\frac{ G_t }{ \hat{E}^2 } \biggr]^{1/2} \biggr\} ~ ,  
 \end{eqnarray} %% {equation}
 where $ X \equiv g_{\phi\phi} G_t (1 - \xi^2 ) / \rho_w^2 $.  
 For $M^2$ not to diverge on the way to the distant region, the condition, 
 $\tilde{L} \Omega_F = Y_{\rm A}   < ( \tilde{L} \Omega_F )_{\rm A}^{+}  < 1$,   
 must be satisfied at the Alfv\'{e}n surface. 
  
% \smallskip 

 Here, the ratio of the toroidal to poloidal magnetic fields is introduced as a model of the 
 magnetic field line shape by the function $\beta^2(r;\Psi)$.  As a simple situation of the 
 poloidal magnetic field, it may be assumed that a magnetic flux tube is distributed along 
 one conical magnetic flux surface $\Psi(\theta_0)=$ constant, where $\theta_0$ is the angle 
 of the magnetic surface to be considered, and it shows a spiral shape on this sheet. 
 Notably, because the functional form of the magnetic surface $\Psi(r,\theta)$ in the entire 
 magnetosphere is not given, the cross-sectional area of the magnetic flux tube along the 
 magnetic flux surface may differ from that of the monopole (or split-monopole)  magnetosphere. 
 In the conventional study of trans-magnetosonic flow solution, one trans-magnetosonic 
 solution was selected from the solution curve group generated by combining field-aligned 
 parameters and give the critical condition at the magnetosonic point. On the other hand, 
 the handling introduced here has the merit that by giving a regular function $\beta^2(r;\Psi)$  
 [ or $\xi^2(r;\Psi)$ ],  a trans-magnetosonic flow can be solved without fine-tuning the critical 
 values of the field-aligned flow parameters at the magnetosonic point. 

%\smallskip 

\subsection{ Constraints on Magnetic Field Lines }

In the following, the function form of $\xi^2(r;\Psi)$  [instead of $\beta^2(r,\theta)$] is assumed 
to be a regular function at the magnetosonic surface, and the magnetized plasma flows 
depending on the field-aligned conserved flow parameters are considered. Although the 
toroidal component of magnetic field $B_\phi(r;\Psi)$ is specified by combining the 
field-aligned parameters, the cross-sectional area change along the flow is also specified 
accordingly. Moreover, the acceleration efficiency and the magnetization parameter in the 
flow are determined depending on the function $\xi^2(r;\Psi)$ and the field-aligned parameters. 
As the functional form of $\xi^2(r,\theta)$, the one considered in TT03 and TT08 is introduced.  

%\smallskip 

From the condition $ A(r,\theta) > 0 $ for the trans-magnetosonic flow solution, we have 
\begin{equation}
     \xi^2 (r,\theta) < ( k + 1 ) \frac{ g_{\phi\phi}( \Omega_F - \omega )^2 \rho_w^2 }
                             { ( G_\phi \hat{E} + G_t \hat{L} )^2 } 
         ~ =   \frac{ -( k + 1) }{ g^{tt} \hat{E}^2 ( 1 - \tilde{L}\Omega_F/Y )^2} ~.  
\end{equation}
At the far distant region ($r/m \gg 1$), this condition becomes 
\begin{equation}
     \xi^2 (r,\theta) < 1 - \frac{1}{ \hat{E}^2 } + \frac{1}{ \hat{E}^2 } \frac{ 2m }{ r } + O( r^{-2} ) ~ .  
\end{equation}
(This corresponds to $ \zeta_0 < (1/\hat{E}^2)(2m/r) $, where $\zeta_0$ is introduced in 
Eq.~(A3) of TT08. ) 
Referencing TT03 and TT08, we consider the following $\xi^2$ model for each type of BH 
magnetosphere. For the outflows of type I/III/IIa, we can apply the following as a functional 
form of $\xi^2(r;\Psi)$,    
\begin{equation}
      \xi^2 =  1 - \frac{\Delta}{\Sigma} \frac{1}{ \hat{E}^2 } +\zeta  ~ .        \label{eq:xi2-1}
\end{equation}
The effect of $\zeta$ is related to the energy conversion efficiency of the outflow, and 
discussed in~\cite{PT20}.  Assuming $\zeta=0$ for simplicity,  the radial velocity by 
Eq.~(\ref{eq:pol_velocity}) becomes 
\begin{equation}
    (u^r)^2 = \frac{ \left[ \, \hat{E}^2 ( 1 - \tilde{L}\Omega_F )^2 - \alpha \, \right] 
    \left[ \hat{E}^2 - (\Delta/\Sigma) \right] }{ \hat{E}^2 \left( \Sigma^2 / {\cal A} \right) - \alpha }  \ . 
\end{equation}
At a distant region ($R \equiv r \sin\theta \gg 1$),  the terminal velocity of the trans-fast 
magnetosonic relativistic jet becomes $(u^r)_\infty = \gamma_\infty \sim \hat{E}$.  
Thus, the total energy is primarily kinetic in the distant region.  On the way to the distant 
region, the initially magnetically dominated MHD flow energy is converted to the kinetic 
energy of the plasma. 

%%\smallskip 

 When we discuss the inflows, certain restrictions on $\xi^2$ should be considered 
 (TT08).  First, it must be $\xi^2=1$  on the event horizon. For example, for type~I or~III 
 inflow case,  we can set up the following functional form: 
\begin{equation}
      \xi^2 =  1 - \frac{\Delta}{\Sigma} f(r, \theta) ~ ,        \label{eq:xi2-1A}
\end{equation}
 where the function $f(r, \theta)$ is a regular function. When the anchor radius appears in the 
 flow, where the toroidal magnetic field becomes $0$, it is necessary to satisfy the condition 
 $\xi^2=\infty$ (or $\beta^2=0)$. 
 Therefore,  we can consider the following as an example,  
\begin{equation}
      \xi^2 =  \left( 1 - \frac{\Delta}{\Sigma} f(r, \theta) \right) 
                  \frac{( G_\phi \hat{E}+G_t \hat{L} )_{\rm H} }{( G_\phi \hat{E}+G_t \hat{L} )}  ~ .        
                  \label{eq:xi2-1B}
\end{equation}
 For type~II, the ``corotation radius''  defined by the radius of $\omega(r;\Psi) = \Omega_F $ 
 occurs in the BH magnetosphere. At the corotation radius where 
 $G_\phi(r,\theta;\Omega_F,a) = 0$, we obtain the finite poloidal velocity [i.e., 
 $(u^r)^2 =$ finite], but we get $M^2 = 0$ and $n = \infty$ where ${\cal B}_p = \infty$ 
 and $\beta^2 = 0$ because $\xi^2(r;\Psi) =$ finite $\neq 0$ is assumed. This indicates the 
 break down of the flow solution there; therefore, we require $\xi^2 = 0$ at the corotation  
 radius.  Thus, we can introduce the following functional form: 
\begin{equation}
      \xi^2 =  \left( 1 - \frac{\Delta}{\Sigma}  f(r, \theta)  \right)  
                  \left( \frac{ \omega - \Omega_F }{ \omega_{\rm H} - \Omega_F } \right)^2  ~ . 
                   \label{eq:xi2-2}
\end{equation}

%%\smallskip  

For type~IIa outflows (discussed in Section~\ref{sec:param_deps}), 
the above solution~(\ref{eq:pol_velocity}) can be applied in the area outside the corotation 
radius, where $G_\phi > 0$.  Fortunately, because the separation surface is located 
outside the corotation radius, there should be no problem with the general outflow, 
starting from the plasma source located near the separation surface. 
However, to obtain physical inflow solution (i.e., type~IIb/IIc) starting around the separation 
surface, we must set up $\xi^2 = 0$ at the corotation radius, hence we should use 
Eq.~(\ref{eq:xi2-2}) as a model of $\xi^2$ rather than Eq.~(\ref{eq:xi2-1A}). 
In the following sections,  we focus only on outflows ejected from outside the corotation 
radius  as the jet model.  

%% \smallskip 

\subsection{ Injection Surface and Initial Velocity }

The outgoing and ingoing plasma flows are injected into the outer and inner BH  
magnetospheres from the plasma source, respectively.  The plasma source must be 
located between the inner and outer light surfaces  so that $\alpha_{\rm inj} > 0$ is  
required.  At the injection surfaces $r = r_{\rm inj}(\theta)$ with $M^2_{\rm inj} \ll 1 $,  
from Eq.~(\ref{eq:poloidal-M2-cold}), we have the relation of 
$\alpha_{\rm inj} = (\hat{E} - \hat{L} \Omega_F)^2$. Because the particle number flux per 
magnetic flux tube $\eta(\Psi)$ is the conserved quantity along a magnetic field line, 
for an injected plasma from the injection surface with a low velocity $u^r_{\rm inj} \ll 1$, 
the number density of flow is very high $ n_{\rm inj} \gg 1$. 
If $\alpha_{\rm inj} > (\hat{E} - \hat{L}\Omega_F)^2 $, we have $(u^r_{\rm inj})^2 < 0$; i.e., 
there is  no physical solution that flows out from the injection surface with a finite initial velocity.   
Specifically, for a highly magnetized flow of $\tilde{L}\Omega_F \sim \hat{E}$, the physical flow 
would start near the light surface, $r_{\rm inj} \lesssim r_{\rm L}$. 
Meanwhile, for a small value of the angular momentum, 
$ |\hat{L}| < ( \hat{E} - \sqrt{\alpha_{\rm inj}} )/ |\Omega_F|$, the flow ejected from the injection 
surface initially has a relativistic speed (i.e., $u^r_{\rm ini} \gtrsim 1$).

%------------------------------------------------------------------------------------------------------------
\begin{figure*}[t] % [htbp]	
\includegraphics[width=17.5cm,clip]{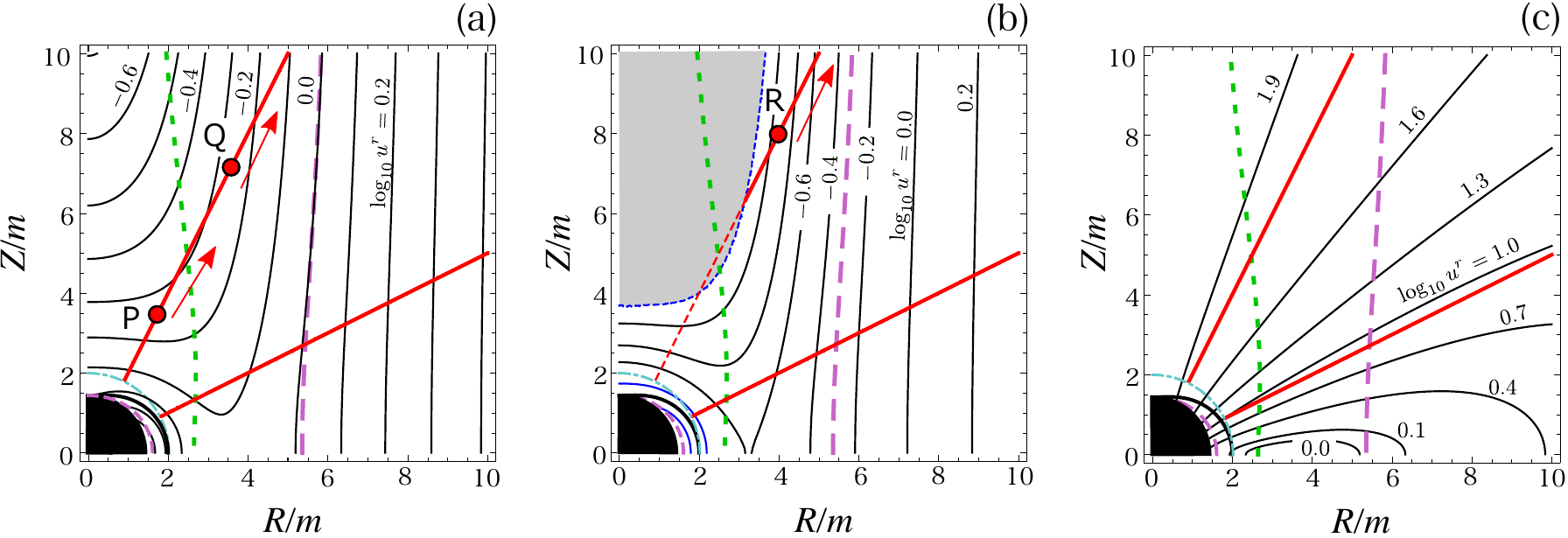}  
\caption{ 
      The distribution of the radial velocity $u^r(R, Z)$ in the BH magnetosphere of 
      $a = 0.9m$, $\Omega_F = 0.5 \omega_{\rm H}$;  
      (a) the cases of $\tilde{L}\Omega_F = 0.9$,  (b) $\tilde{L}\Omega_F = 0.93$,  
      and  (c) $\tilde{L}(\Psi) = \tilde{L}_0 \sin^2\theta$, where $\tilde{L}_0 \Omega_F = 0.9$. 
      The broken magenta curves show the locations of the inner and outer light surfaces. 
      The dotted green curve shows the location of the separation surface. 
      The shaded area on the upper left of the dotted curve in (b) is the forbidden region, 
      where $u_p^2 < 0$. 
      The red lines (outside the corotation surface of the cyan dashed curve)  
      are examples of the streamlines for outflow ($Z=2R$, $Z=0.5R$). 
      The red dotted line in (b) is unacceptable as an outflow.  The black region shows 
      the BH.  Points {\sf P}, {\sf Q}, and {\sf R} are examples of the injection points.  
     }
 \label{fig:RZ-Ur} %%-- Figure_2  
\end{figure*}
%------------------------------------------------------------------------------------------------------------//

%%\medskip 

\section{ Parameter Dependencies on MHD Outflow }    \label{sec:param_deps}

We consider the parameter dependence of outflow in detail.  
For numerical plots of the radial velocity $u^r(r;\Psi)$ about outflows, 
we set up the functional form of $\xi^2(r;\Psi)$.  Hence, from Eq.~(\ref{eq:poloidal-M2-pm}), 
we obtain the distribution $M^2(r;\Psi)$ along the flow line; using Eq.~(\ref{eq:Bf}), we have 
the toroidal component of the magnetic field ${\cal B}_\phi(r;\Psi)$ and  ${\cal B}_p(r;\Psi)$. 
As such, assuming the function $\xi^2(r,\theta)$ is equivalent to assuming the poloidal 
magnetic field ${\cal B}_p$.  We discuss trans-fast magnetosonic flow solutions along a single 
flux tube that is theoretically and observationally valid so that we can consider the qualitative 
understanding of the behavior of jet flow acceleration. 

%%\smallskip 

 Certain physical parameters would be significantly affected by the general-relativistic effect; 
 hence, we focus our research on this. The locations of the light and Alfv\'{e}n surfaces,   
 which are the typical scales of the magnetosphere, primarily depend on the angular velocity 
 $\Omega_F$ of the magnetic field lines and the ratio of the rotational energy to the total 
 energy $\tilde{L} \Omega_F$ of the plasma flow, respectively.  In the following parameter 
 search, the dependency of $\Omega_F$ and $\tilde{L}$ is particularly examined. 
 We assume $\Omega_F(\Psi) =$ constant for the entire magnetosphere. Although an 
 example  of $\Omega_F(\Psi) =$ non-constant is considered by Pu \& Takahashi \cite{PT20}, 
 the basic properties are similar, whereas the distributions of the light surfaces differ slightly. 
 Moreover, we assume $\eta(\Psi) =$ constant for the entire magnetosphere for simplicity. 
 Notably, in the $\xi^2$ model, the parameter $\eta$ is renormalized into the magnetic field 
 ${\cal B}_p$ and ${\cal B}_\phi$ so that it does not directly appear in the expression of the  
 MHD flow equations. 
 
%%\smallskip 
 
Figure~\ref{fig:RZ-Ur} shows the velocity distribution $u^r(R, Z)$       
when Eq.~(\ref{eq:xi2-1})  is employed as the distribution of $\xi^2(R, Z)$  
with constant $\hat{E}$ and $\Omega_F$ for a rapidly rotating BH (type~IIa) case, 
where $R = r \sin\theta$ and $Z \equiv r \cos\theta$.  To understand the angular momentum 
dependence on the flow, the cases of constant angular momentum $\hat{L}(\theta)=$constant 
are plotted in Figures~\ref{fig:RZ-Ur}(a) and~\ref{fig:RZ-Ur}(b), and the case where $\hat{L}$ 
has angular dependence of $\hat{L}(\theta) = \hat{L}_0 \sin^2\theta$ with a constant $\hat{L}_0$ 
is plotted in Figure~\ref{fig:RZ-Ur}(c).  The outflow from the plasma source (e.g., points {\sf P}, 
{\sf Q}, and {\sf R} in Fig.~\ref{fig:RZ-Ur}) flows out in the distant region. 

%%\smallskip 
 
If $\hat{L}(\theta) =$ constant in the magnetosphere, the higher the latitude, the smaller the 
initial velocity. However, when the angular momentum is very large, the flow's forbidden region 
(i.e., $u_p^2<0$ region) appears in the high latitude region [see, the shaded region in 
Fig.~\ref{fig:RZ-Ur}(b)].  Alternatively, the angular momentum at high latitudes should be 
selected not too large for the trans-fast magnetosonic solution to fill the magnetosphere in 
the jet region [see, Fig.~\ref{fig:RZ-Ur}(a)].  Moreover, we observe that the behavior of 
acceleration differs depending on whether the angular momentum is constant throughout 
the magnetosphere or whether it has a $\sin^2\theta$ dependence  [see, Fig.~\ref{fig:RZ-Ur}(c)]. 
To fit the acceleration profile of the M87 jet (see Fig.~\ref{fig:M87-KAVA}),  
$\tilde{L}\Omega_F =$ constant seems to be better in the entire region of the magnetosphere.  
Afterward, we will consider the case of $\hat{L}(\Psi)=$ constant. 

%%\smallskip 

Figure~\ref{fig:RZ-Ur} also shows the velocity distribution within the separation surface, 
where the parameter set used here corresponds to the type~IIa case; therefore, it fails 
as a physical solution on the corotation surface. 
Notably, from the area between the corotation and separation surfaces (e.g., point {\sf P}),  
an outgoing flow ejected with a sufficiently large initial velocity is possible.  

%%\smallskip 

%------------------------------------------------------------------------------------------------------------
\begin{figure}[h] % [htbp]
\includegraphics[width=8cm,clip]{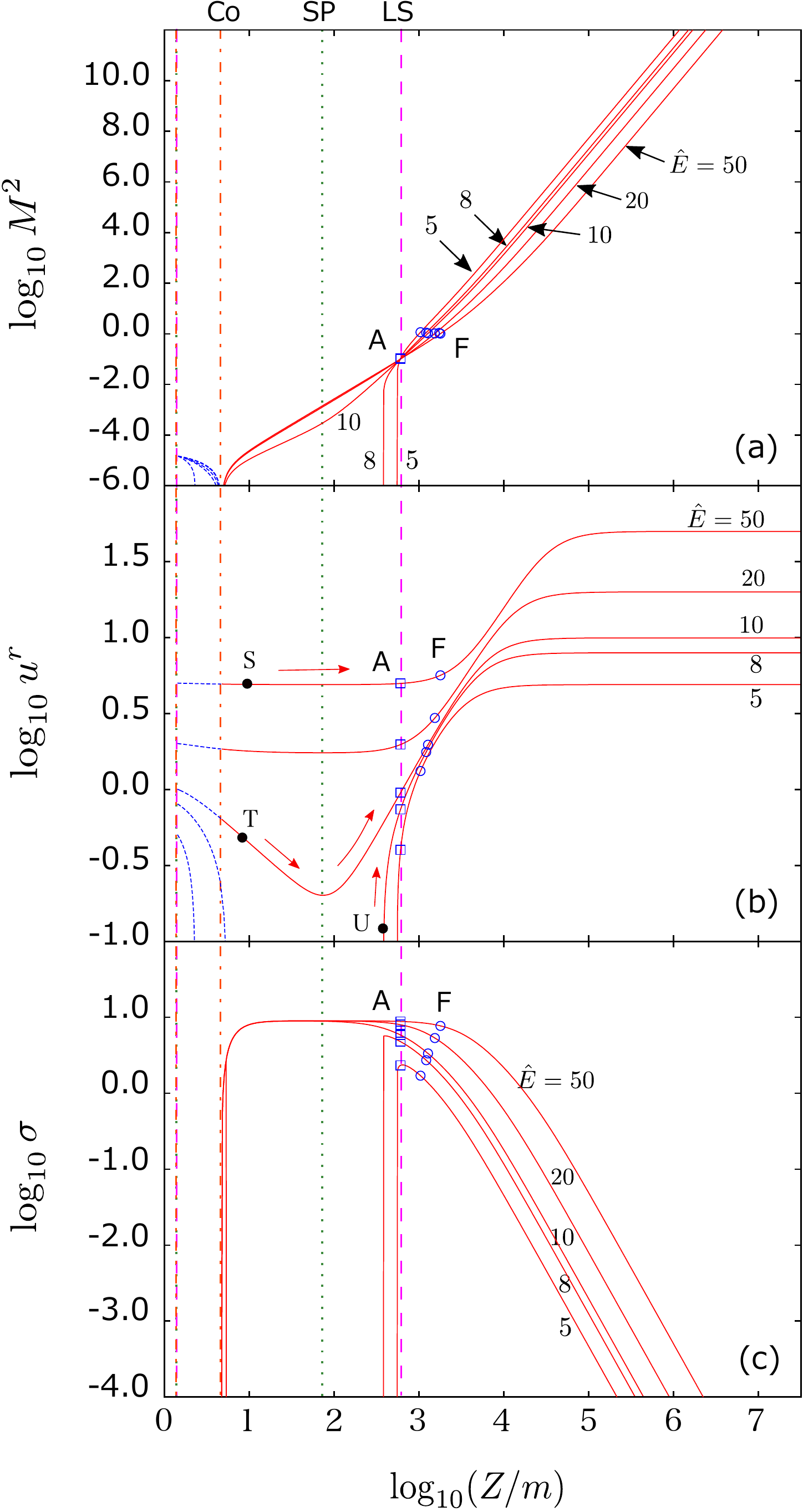}  
\caption{  %%%=== {\bf Energy $E$ dependence ---}
    The dependences of total energy $\hat{E}$ on relativistic outflows
    (Type~IIa: $\hat{E}=\{5.0, 8.0, 10.0, 20.0, 50.0 \}$ with $a=0.9m$, 
    $\Omega_F=0.05 \omega_{\rm H} = 0.0157/m$, $\tilde{L}\Omega_F=0.9$, 
    $\theta=1/\hat{E}$) 
    (a) the Alfv\'{e}n Mach number $M^2(Z)$, (b) the radial velocity $u^r(Z)$ and 
    (c) the magnetization parameter $\sigma(Z)$.   
    The blue curves inside the corotation surface ({\sf Co}) are unphysical solutions 
    as outflows. 
    }
 \label{fig:energy-dependence}    %%-- Figure_3 
\end{figure}
%------------------------------------------------------------------------------------------------------------
\begin{figure}[htbp]
\includegraphics[width=8cm,clip]{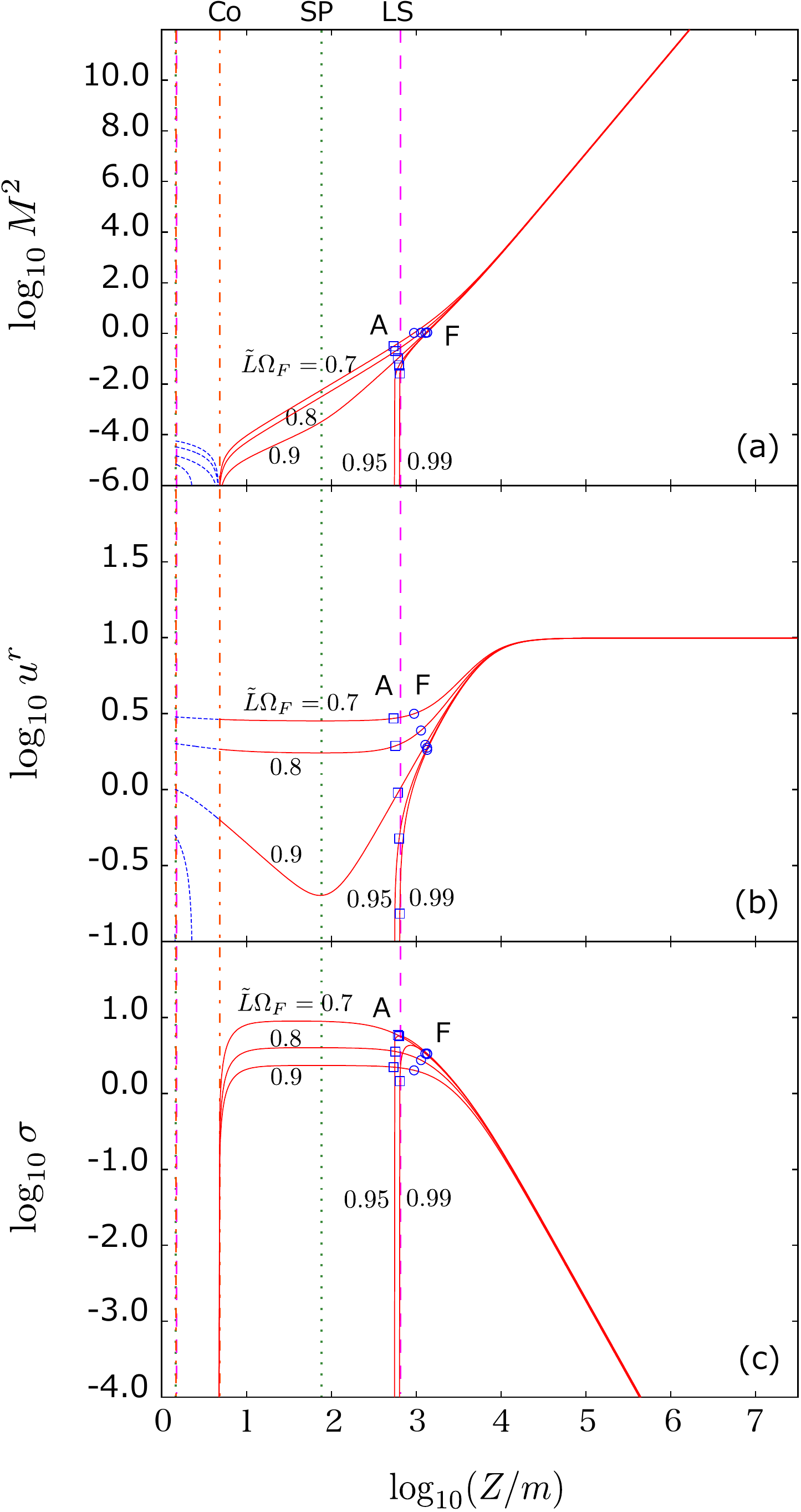}  
\caption{  %%%===  {\bf Angular Momentum $L$ dependence ---}
    The dependences of total angular momentum $\hat{L}$ on relativistic outflows 
    (Type~IIa: $\tilde{L}\Omega_F = \{ 0.70,~ 0.80,~ 0.90,~ 0.95,~ 0.99 \}$, $a=0.9m$, 
    $\Omega_F=0.05 \omega_{\rm H} = 0.0157/m$, $\hat{E}=10.0$, $\theta=1/\hat{E}$) 
    (a) the Alfv\'{e}n Mach number $M^2(Z)$, (b) the radial velocity $u^r(Z)$ and 
    (c) the magnetization parameter $\sigma(Z)$.         
  }
 \label{fig:ang_mom-dependence}   %%-- Figure_4 
\end{figure}
%------------------------------------------------------------------------------------------------------------
\begin{figure}[htbp]  %%[t]
\includegraphics[width=8cm,clip]{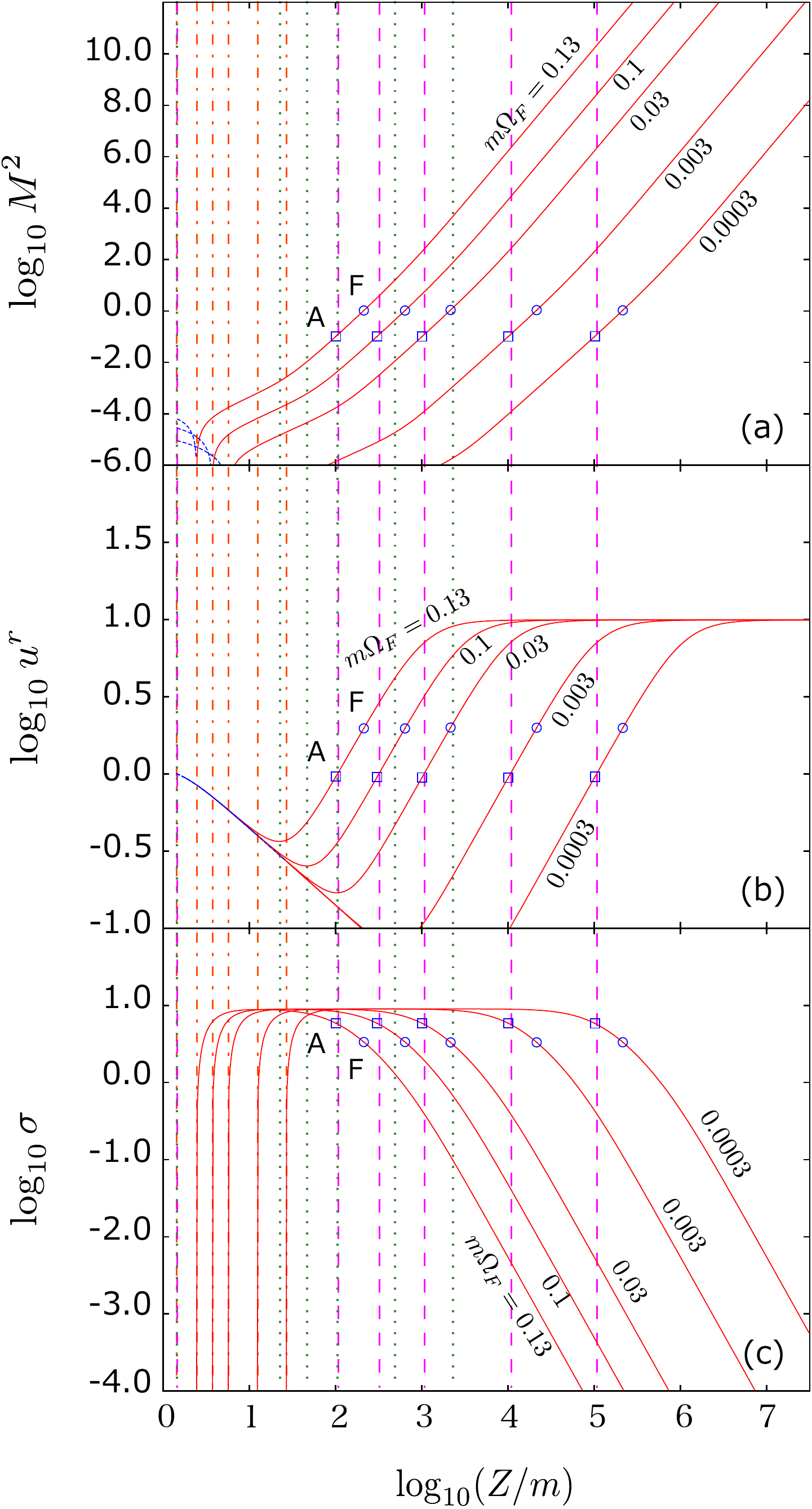}  
\caption{   %%%==={\bf Angular Velocity $\Omega_F$ dependence ---}
    The dependences of magnetosphere's angular velocity $\Omega_F$ on relativistic outflows 
    (Type~IIa: $m\Omega_F = \{ 0.0003,~ 0.003,~ 0.03,~ 0.10,~ 0.30 \}$, $a=0.9m$,  
     $\hat{E}=10.0$, $\tilde{L}\Omega_F=0.9$, $\theta=1/\hat{E}$) 
    (a) the Alfv\'{e}n Mach number $M^2(Z)$, (b) the radial velocity $u^r(Z)$ and 
    (c) the magnetization parameter $\sigma(Z)$.     
 }
 \label{fig:omega_F-dependence}      %%-- Figure_5 
\end{figure}
%--------------------------------------------------------------------------------------------------------------//

 Figures~\ref{fig:energy-dependence}--\ref{fig:omega_F-dependence} show the dependences  
 of the field-aligned parameters on the trans-fast magnetosonic outflow solutions (red curves)  
 along a  $\theta=$ constant magnetic field line for the type~IIa case.  
 The label ``{\sf Co}'' or orange dashed line indicates the location of the corotation surface.  
 The label ``{\sf SP}'' or green dotted line indicates the location of the separation surface and   
 the label ``{\sf LS}'' or magenta broken lines indicate the location of the light surfaces.  
 The mark $\square$ labeled  ``{\sf A}'' and the mark {\Large $\circ$} labeled ``{\sf F}'' 
 indicate the Alfv\'{e}n and fast-magnetosonic points, respectively. 

%%\smallskip 

\subsection{ Energy $E$ Dependence }

 In Figure~\ref{fig:energy-dependence}, we show the $\hat{E}$ dependence on the trans-fast 
 magnetosonic outflow solutions,   
 where the values of $a$, $\tilde{L}\Omega_F$, and $\Omega_F$ are fixed; therefore, the 
 locations of the separation surface, light surfaces, and Alfv\'{e}n surfaces  are the same.   
 The Alfv\'{e}n Mach number $M^2(Z)$ rapidly increases from around the Alfv\'{e}n 
 point and passes through the fast-magnetosonic point to a distant region. 
 For large $\hat{E}$ values, where the $\hat{L}$ value is also large, the location of the 
 fast-magnetosonic point appears on the outside,  and the plot of the radial velocity $u^r(Z)$ 
 shifts upward over the entire area. 
 In Figure~\ref{fig:energy-dependence}(b), the injected plasma around the separation surface 
 has a nonzero initial velocity. For example, the outflow solution passing through point {\sf S} 
 has an almost constant flow velocity between the corotation and Alfv\'{e}n surfaces  
 and then  accelerates across the fast-magnetosonic surface.  
 For an outflow solution passing through point {\sf T}, the outflow decelerates once 
 toward the separation surface, but after passing the separation surface, it accelerates and 
 passes through the Alfv\'{e}n and fast-magnetosonic surfaces in order. 
 For a low energy flow, the plasma flows out at a very low speed from the vicinity of the 
 Alfv\'{e}n surface (point {\sf U}), which is far outside the separation surface.   
 Such outflow is significantly accelerated beyond the Alfv\'{e}n surface from the injection surface.  

%%\smallskip 

 The rescaled number density $\tilde{n} \equiv n / (4\pi\mu_{c}\eta^2)  = 1/M^2$ decreases 
 in the distance,  unlike the increase in the Alfv\'{e}n Mach number 
 [see, Fig.~\ref{fig:energy-dependence}(a)].  Figure~\ref{fig:energy-dependence}(c) shows 
 the distribution of the magnetization parameter $\sigma(Z)$. 
 For a large energy outflow,  the magnetic field energy is dominant at the beginning of the 
 flow; however, as the  outgoing fluid accelerates, the fluid part of the energy increases and 
 becomes dominant at a far distance ($\sigma \ll 1$). 
 For a low energy outflow, which corresponds to a large angular momentum outflow, 
 the fluid part of the energy is dominant at the beginning of the flow when the injection point 
 is located just inside the outer light surface.  

 For a slowly rotating BH case of type~I, both the outer light surface and the outer Alfv\'{e}n 
 surface shift outward; however,  the asymptotic feature of outflows is the same. 

%%\smallskip 

\subsection{ Angular Momentum $\tilde{L}\Omega_F$ ($L$) Dependence }

 Figure~\ref{fig:ang_mom-dependence} shows the $\tilde{L}\Omega_F$-dependence on 
 the trans-fast magnetosonic outflow solutions, where $\hat{E}$ and $\Omega_F$ are fixed; 
 i.e., the dependence of angular momentum $\hat{L}$ is investigated. 
 The angular momentum $\hat{L}$ determines the behavior of the initial stage of acceleration 
 of the ejected outflow in the sub-Alfv\'{e}nic region and near the light surface. 
 For outflows with small angular momentum, the velocity distribution in the sub-Alfv\'{e}nic 
 region has a large value, i.e., a solution with nonzero velocity is obtained around the 
 separation region.  Meanwhile, for outflows with large angular momentum, 
 $\tilde{L}\Omega_F \sim 1$,  no physical solution region appears around the separation 
 surface in the sub-Alfv\'{e}nic region. 
 Such a flow with large angular momentum starts at a slower speed from slightly inside 
 the outer light surface, where the toroidal motion of the flow is dominated;  i.e., 
 $v^\phi_{\rm ini} \approx c$ and $\gamma_{\rm ini} \gg 1$ (see also, \cite{TS98}).  
 Thus, the value of $\hat{L}$ is related to the toroidal motion in the sub-Alfv\'{e}nic region.  
 The closer the injection surface to the light surface, the more dominant toroidal motion 
 of plasma ($\sigma_{\rm ini} \ll 1$) is obtained in the initial stage of the MHD flow, 
 whereas the poloidal motion is relatively small.  Subsequently, it is converted to poloidal 
 motion, and eventually becomes magnetically dominated MHD outflow ($\sigma \gg 1$) 
 in the next stage of the acceleration region. Through the equipartition region (TT03), 
 the kinetic energy by the poloidal motion becomes dominant over the magnetic energy;  
 i.e., $\sigma_\infty \ll 1 $.    
 The terminal velocity $(u^r)_\infty$ does not depend on angular momentum $\hat{L}$.  
 When the angular momentum becomes large, the flow starting near the separation 
 surface is prohibited. 
  
%%\smallskip 

\subsection{ Angular Velocity $\Omega_F$ Dependence }
           
 Figure~\ref{fig:omega_F-dependence} shows the $\Omega_F$-dependence on the 
 trans-fast magnetosonic outflow solutions, where $\hat{E}$ and $\tilde{L}\Omega_F$ are 
 fixed; therefore, a large (or small) $\Omega_F$ value corresponds to a small (or large) 
 $\hat{L}$ value.  Each solution curve in Fig.~\ref{fig:omega_F-dependence} is similar,  
 although the location of the light surface depends on both $\Omega_F$ and $a$. 
 When the value of $\Omega_F$ decreases, the outer light surface shifts to the outside. 
 Meanwhile, when the value of spin $a$ increases, the effect of the spacetime 
 dragging on the magnetic field lines becomes large and it shifts to the inside. 
 
%% \medskip  

\section{  Plasma acceleration on M87 jet  } %================================

 Now, we explain the observed data of the M87 jet by parameter searching 
 using the stationary trans-fast magnetosonic outflow model.  The mass of the central BH 
 in M87 is estimated to be $m_{\rm M87} = ( 6.5 \pm 0.7) \times 10^{9} m_\odot$  
 (e.g., \cite{Gebhardt+11,EHT+19a}).        
 The inclination angle (viewing angle) of the M87 jet is estimated to be  
 $\theta_{\rm inc} \approx 17^\circ$ \cite{Walker+18}.  
 The jet power in M87 is in a range of $10^{35}$--$10^{37}$  J s$^{-1}$, and there are 
 several theoretical calculations of jet power driven by electromagnetic mechanisms 
 represented by the BZ process \cite{BZ77} and/or Blandford--Payne process 
 \cite{BP82} (see also, e.g.,  \cite{Li+09}). 
 The parsec-scale jet of the galaxy M87 is roughly parabolic ($Z\propto R^{1.7}$) in structure 
 \cite{AN12,NA13}, where $Z$ is the deprojected distance along the jet, and $R$ is the jet 
 width, and the M87 jet becomes conical-like ($Z\propto R^{1.3}$) near the BH \cite{Hada+13}.  
 The radial profile of the M87 jet velocity is observed by 
 \cite{Kovalev+07, AN13, Mertens+16, Hada+16, Hada+17, Walker+18}, 
 and recently detail data for the region closer to the base of the jet has been obtained by 
 Park et al. \cite{Park+19} using the KaVA.  

%%\smakkskip 

\subsection{ Acceleration Region of M87 Jet }

 In this study, we examined the general-relativistic effects in the outer BH magnetosphere. 
 In this section, the M87 jet data fitting is performed using the general-relativistic version of 
 the $\xi^2$ model by TT08 that incorporates the effect of a rotating BH,  where the plausible 
 magnetic field suggested by VLBI observations and general-relativistic MHD (GRMHD) numerical 
 simulations for the relativistic jet are adopted (e.g., \cite{MCA18,McKinney06,Nakamura+18}).  
 To discuss the acceleration properties of the jet,  we can consider the angular momentum 
 dependence on the outflow in the acceleration region and the initial velocity around the plasma 
 source region. 
 
%%\smakkskip 
 
\begin{figure*}[htbp]
\includegraphics[width=7.8cm,clip]{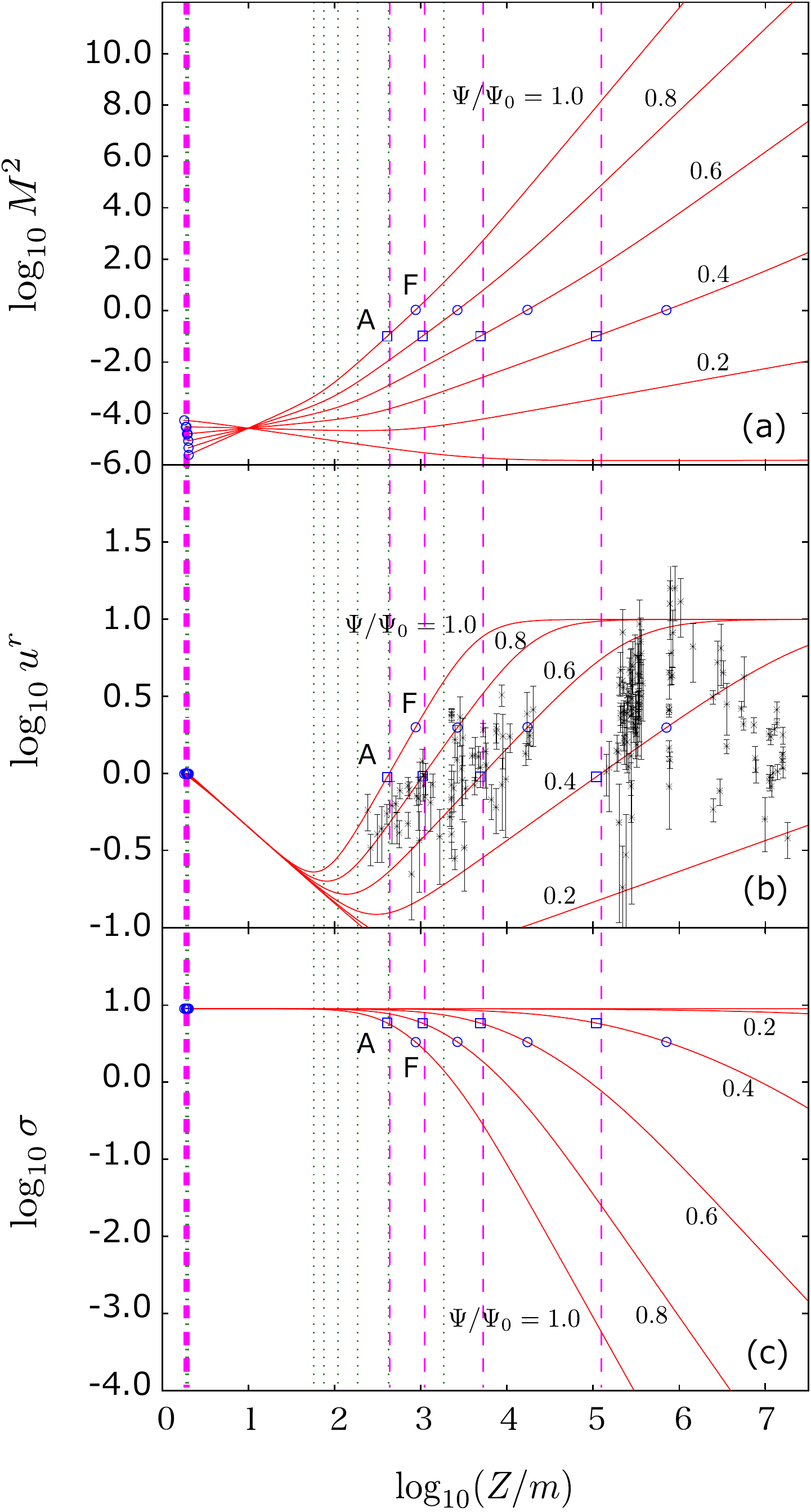}  
\hspace{3mm}
\includegraphics[width=7.8cm,clip]{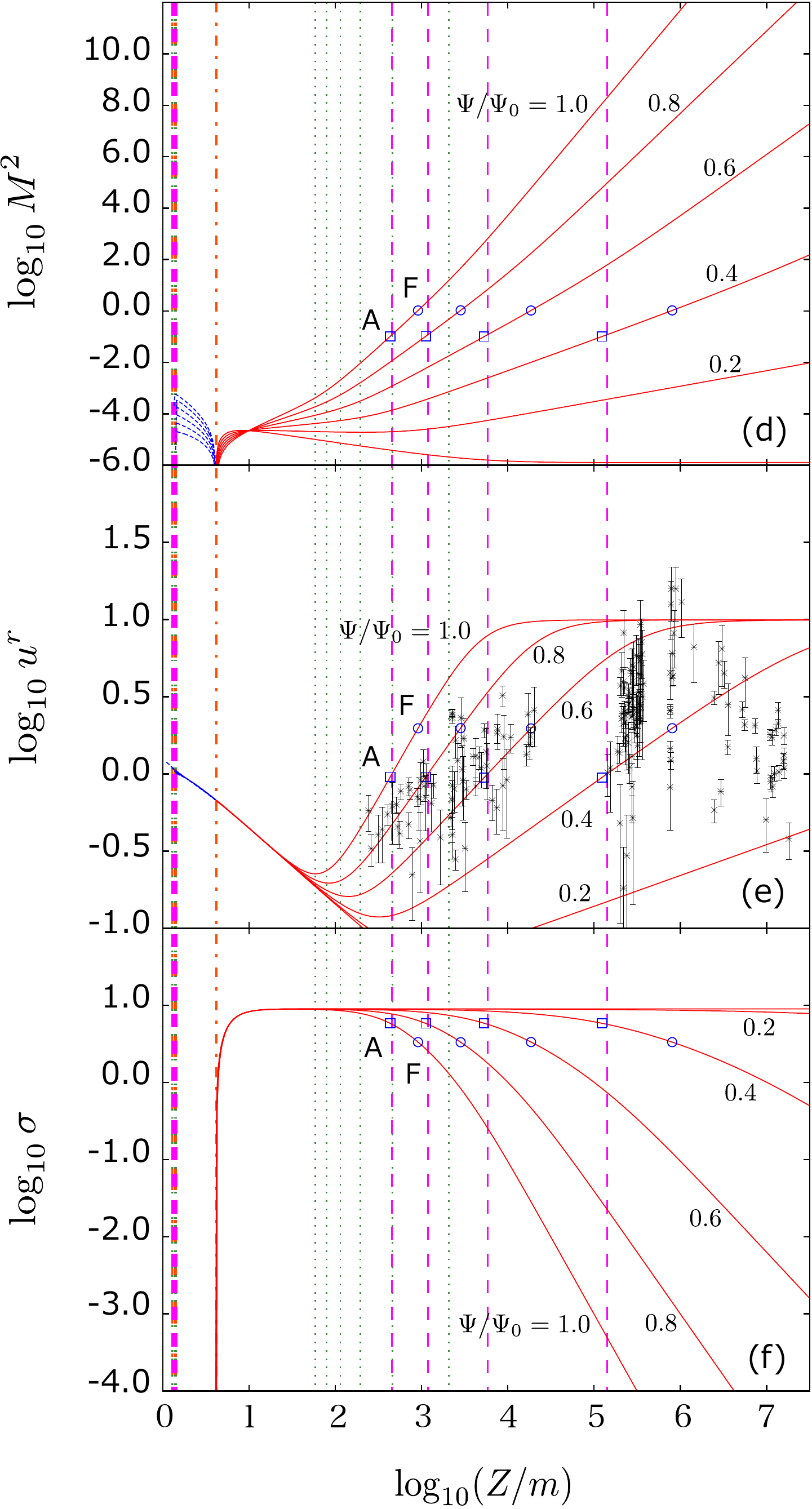}  
\caption{ 
    The $\Psi$-dependence on the jet solution is shown 
    ($ \Psi/\Psi_0 = \{ 1.0,\, 0.8,\, 0.6,\, 0.4,\, 0.2,\, 10^{-5} \} $, $\theta_0=1/\hat{E}$).   
     (a)--(c) type~I:  a slowly rotating BH of $a = 0$ with $\hat{E} = 10.0$, 
     $\Omega_F = \omega_{\rm H} + 0.12 (\Omega_{\rm max}-\omega_{\rm H}) = 0.0231/m$, 
     $\tilde{L}\Omega_F = 0.9$,  and 
     (d)--(f) type~IIa:  a rapidly rotating BH of $a = 0.9m$ with $\hat{E} = 10$, 
     $\Omega_F = 0.07\omega_{\rm H} = 0.0219/m$, $\tilde{L}\Omega_F = 0.9$ are presented.  
     The measured data points with error bars are adopted from \cite{Park+19}.       
     In panels (d)(e), the dotted blue curves inside the corotation surface 
     show unphysical branches as the jet solution. 
 }
 \label{fig:M87-KAVA}             %%-- Figure_6 
\end{figure*}

 In the following, we apply the trans-fast magnetosonic jet solution to the Park's M87 data 
 \cite{Park+19}. As the flow's streamlines in the BH magnetosphere, we apply the approximated 
 magnetic field configuration in the asymptotic region obtained from Eq.~(57) of TT03;  i.e., 
\begin{equation}
  \theta_0 Z/m = (R/m)^{\Psi_0/\Psi}               \label{eq:TT03asymptotic} 
\end{equation}
 for $\theta \ll 1$,  where $\Psi \to 0$ for $\theta \to 0$, although this solution should not extend 
 inside the outer light surface because it is out of the range of the approximation of TT03.  
 However, it may be suitable for around the outer light surface, where current observation 
 data are primarily obtained. We expect it to be applicable for qualitative understanding of 
 the relativistic jets from the sub-Alfv\'{e}nic region of the outer BH magnetosphere.  
 In Figure~\ref{fig:M87-KAVA}, we interpolate the approximated magnetic field shape 
 (\ref{eq:TT03asymptotic})  into the sub-Alfv\'{e}nic region that includes the plasma source,  
 where the boundary layer $\Psi_0 = \Psi(\theta_0)$ between the jet region and middle/lower 
 latitude accreting matter region is given by the $\theta_0 \sim 1/\hat{E}$ conical magnetic 
 field line. Notably, around or inside the outer light surface, Hada et al. \cite{Hada+13} and 
 TT03 model suggests a conical-like magnetic field shape. 
(The nature of the conical outflow has been discussed in Section \ref{sec:param_deps}.) 

%%\smallskip 

%-------------------------------------------------------------------------------------------------------m87---
%-------------------------------------------------------------------------------------------------------m87---

 %%\medskip 
 
 Figures~\ref{fig:M87-KAVA}(b) and~\ref{fig:M87-KAVA}(e) show the radial velocity $u^r(Z)$ 
 of the outgoing jet from the BH magnetosphere.    
 The KaVA's data is rescaled in the unit of $m_{\rm M87}$,  where  
 $1 \, {\rm mas}  \approx 260 \, m_{\rm M87}$ \cite{Park+19}.  
 The correspondence between the apparent angle and actual scale significantly depends 
 on  the inclination angle (viewing angle) and opening angle of the jet. 
 The ejected outflow from the plasma source is accelerated around the light surface; after 
 passing the fast-magnetosonic surface, the flow velocity reaches a nearly constant value 
 that is large enough; i.e., the initially magnetically dominated outflow becomes the fluid's 
 kinetic energy dominated flow.  
 Although we assumed that $a=0.9m$, the spin dependence on the outflow solutions 
is weak at least outside the outer light surface. The efficiency of acceleration differs for each 
magnetic field line. The magnetic field lines close to the funnel wall ($\Psi \sim \Psi_0$) cause 
large acceleration;  moreover, the acceleration efficiency is minute for the magnetic field lines 
near the axis ($\Psi \ll \Psi_0$). The plasma source is located near the separation surface of 
each curve for the magnetic flux surface.  
 
%% \smallskip 
 
 Now, we understand  that  the values of $\hat{E}$ and $\tilde{L}\Omega_F$ are estimated 
 by fitting the terminal velocity $u^r_\infty$ and the initial velocity $u^r_{\rm ini}$ at the injection 
 region (or the acceleration region near the plasma source), and the location of the light surface 
 $r_{\rm L}$  is specified by parameters $\Omega_F$ and $a$.   
 The trans-fast magnetosonic MHD outflow becomes a fluid-dominated outflow at the distant 
 region.  By observing the terminal velocity $u^r_\infty $ (or $\gamma_\infty$) of the M87 jet,  
 we can specify the value of  $\hat{E} \approx 10$.   
 Angular momentum $\hat{L}$ (or $\tilde{L}\Omega_F$) is largely related to $u^r_{\rm ini}$ 
 and the velocity distribution in the acceleration region around the outer light surface.  
 By comparing the theoretical solution with the M87 observed data, we estimate 
 $\tilde{L}\Omega_F \approx 0.9$ from the acceleration profile of the jet. If the angular 
 momentum parameter of the flow reduces by some extent, the ejected flow around the 
 separation surface will have a faster initial velocity. The initial flow acceleration obtained 
 in certain GRMHD simulations (e.g.,  \cite{McKinney06, Nakamura+18, Chatterjee+19})  
 that show $u^r_{\rm ini} \gtrsim 1$ and do not explain the M87 observations well would be 
 explained in a parameter range with a small angular momentum. 

%% \smallskip 

 The location of the outer light surface depends on both the angular velocity of the magnetic 
 field lines and the BH spin.  In Figures~\ref{fig:M87-KAVA}(b) and~\ref{fig:M87-KAVA}(e), 
 from the location of the acceleration region for the M87 jet,  we estimate that  
 $\Omega_F \approx 0.023/m$ for $a = 0$ case, and 
 $\Omega_F \approx 0.022/m$ for $a = 0.9m$ case.   
 Thus, the outer light surface, $R_{\rm L} \sim c/\Omega_F$, will be located far enough from 
 the central BH; i.e., the effect of BH spin decreases at the outer light surface. 
 It can be said that the behavior of the jet does not depend on the details of type~I or type~IIa. 
 
%% \smallskip 

\subsection{ Energy Conversion in M87 Jet }

 Figures~\ref{fig:M87-KAVA}(c) and~\ref{fig:M87-KAVA}(f) show the distribution of the 
 magnetization parameter $\sigma(Z)$ in the jet. The efficiency of energy conversion 
 from the magnetic energy to the fluid's kinetic energy differs for each magnetic field line. 
 Although the jet has a large value of $\sigma$ at the beginning, the value of $\sigma$ 
 decreases along the magnetic field lines of $\Psi \sim \Psi_0$; then, at a distant region,  
 the jet becomes a kinetic energy-dominated outflow.  
 However, along the magnetic field lines near the axis of rotation, $\Psi \ll \Psi_0$, where 
 the centrifugal force on plasma does not work effectively, the acceleration efficiency is 
 relatively small, and the value of $\sigma$ does not decrease significantly. Thus, the 
 magnetic field lines close to the funnel wall cause efficient energy conversion, whereas 
 the efficiency is minute for the magnetic field lines near the axis.   

%% \smallskip  

 The strength of the magnetic field at the footpoint of the jet is estimated from the Event 
 Horizon Telescope (EHT) observation \cite{Kino+14,Kino+15}.  The value of the magnetization 
 parameter $\sigma$ within the outer light surface is estimated by Kino et al. \cite{Kino+21}; 
 i.e., $ 1 \times 10^{-5} \leq \sigma \leq 6 \times 10^3$ within the radio core with VLBA at 
 43 GHz and $ 5 \leq \sigma \leq 1 \times 10^6$ within the putative synchrotron self-absorption 
 (SSA) thick region in the EHT emission region at 230 GHz. 
 Thus, we identify that the high values of the $\sigma$ profile within the outer light surface 
 agree the $\xi^2$ model for the GRMHD jet.   

%% \smallskip    

Moreover, low-$\sigma $ ($\sigma \sim 10^{-4}$) has been observed in the major atmospheric 
gamma imaging Cerenkov telescope (MAGIC) TeV-gamma observation \cite{Acciari+20}, 
although the accuracy of the radiation source position observation is uncertain. For example, 
in Figure~\ref{fig:M87-KAVA}(c), such a low-$\sigma$ flow can be achieved at a distance of 
$Z/m>10^{6}$. Alternatively, the MAGIC observation region may be near the outer light surface, 
which is because  the outflow solution emitted from the region slightly inside the outer light 
surface is realized with very low-$\sigma_{\rm ini}$ for $\tilde {L} \Omega_F \sim 1 $ 
[see, Figs.~\ref{fig:energy-dependence}(c), \ref{fig:ang_mom-dependence}(c)].  
Such an outflow initially rotates in the toroidal direction at very high speed 
($\gamma_ {\rm ini} \gg 1$), resulting in low-$\sigma$, and then it accelerates outward. 
Interestingly, the plasma moving at high speed in the toroidal direction just inside the outer 
light surface generates an extremely large acceleration, which would be a region where 
very-high-energy gamma rays are generated.
If the vicinity of the outer light surface is the source of high-energy $\gamma$-ray for some 
magnetic field lines in the jet, the $\gamma$-ray source may also be distributed along the jet  
(i.e., hollow cylindrical-shape) because the outer light surface is distributed in the region 
extending along the jet.  

%%\medskip  

\section{ Discussion }

 \subsection{ Dependence on magnetic field line shapes } 

 To discuss the parameter dependence of the ideal MHD field-aligned quantities, we assume 
 the magnetic surface has a conical shape. Further, it is easy to extend to other plausible 
 magnetic field shapes; e.g., we can use the model  
 \begin{equation}
     \theta_0 Z/m = (R/m)^{p(\Psi_0/\Psi)},          \label{eq:B_shape_p}
 \end{equation}
 where $p$ is a parameter for a magnetic field configuration, and $\Psi=\Psi_0$ is the 
 streamline for the jet's boundary wall. 
 According to Hada's observation \cite{Hada+13}, the M87 jet shape near the jet source region 
 bends around the fast-magnetosonic surface, suggesting a conical-like shape on the side 
 near the central BH. This shape would suggest the magnetic field configuration in the BH 
 magnetosphere. When applied to the above model (\ref{eq:B_shape_p}), it is about 
$p \approx 1.3$ for the boundary wall of the funnel on the inside of the bend and about 
$p \approx 1.7$ on the outside.

%%\smallskip
%------------------------------------------------------------------------------------------------------------//
\begin{figure}[htbp]
 \includegraphics[width=8cm,clip]{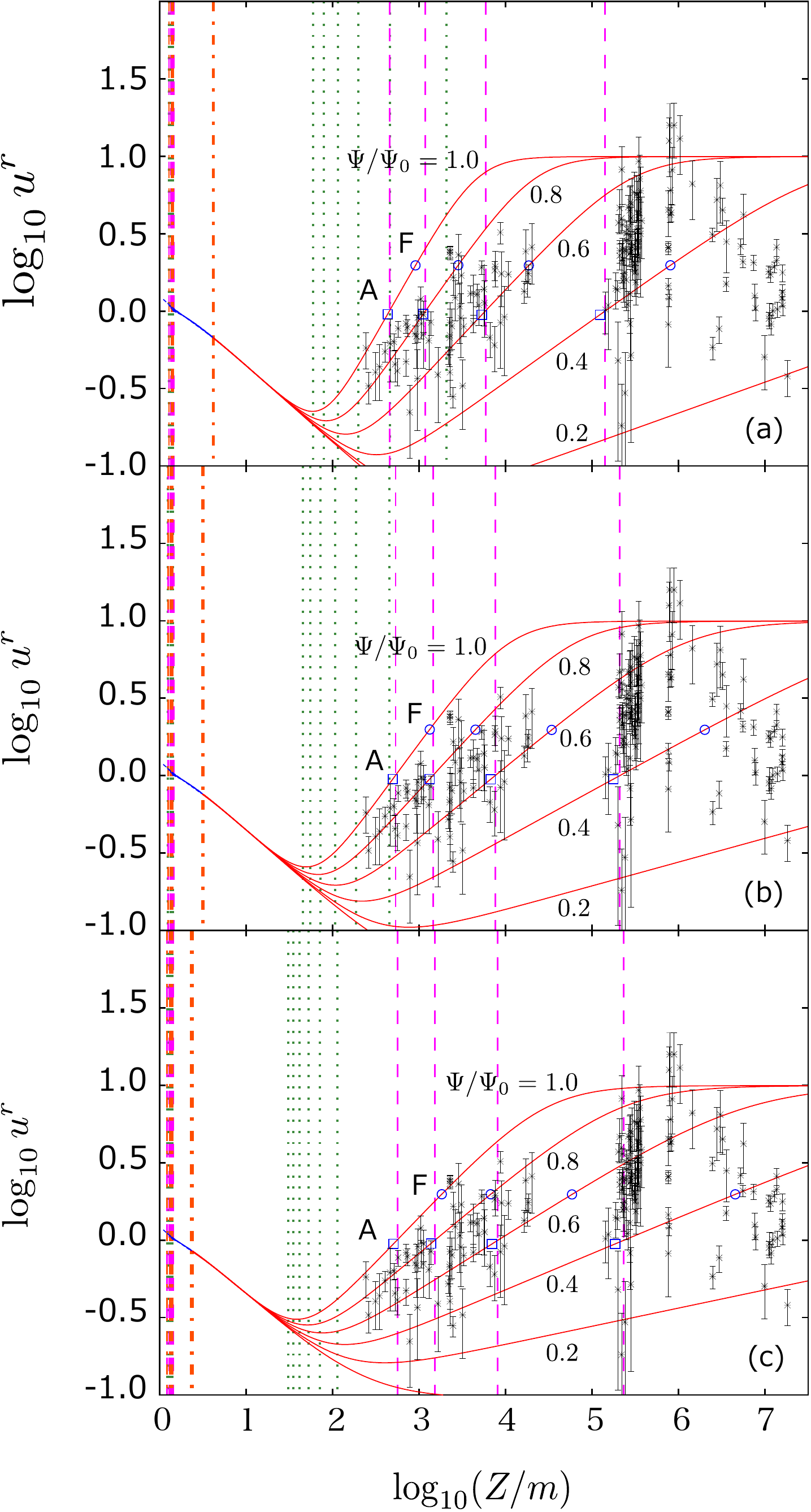} 
 \caption{ 
      The radial 4-velocity $u^r$ of the flow with different magnetic field shapes,  
      (a) $p=1.0$, (b) $p=1.3$, and (c) $p=1.7$, overlapped with the observation jet velocity 
      of M87.  The trans-fast magnetosonic solutions along the magnetic flux surfaces of 
     $(\Psi/\Psi_0) =\{ 1.0,\, 0.8,\, 0.6,\, 0.4,\, 0.2,\, 10^{-5} \} $ are plotted,  where 
     $a=0.9m$, $\hat{E}=10.0$, $\tilde{L}\Omega_F=0.9$, $\theta_0=1/\hat{E}$ and 
     (a) $\Omega_F=0.07 \omega_{\rm H} = 0.0219/m$, 
     (b) $\Omega_F=0.15 \omega_{\rm H} = 0.0470/m$,      
     (c) $\Omega_F=0.30 \omega_{\rm H} = 0.0940/m$ are assumed.    
}
\label{fig:M87-mag}             %%-- Figure_7 
\end{figure}
%------------------------------------------------------------------------------------------------------------//

 Figure~\ref{fig:M87-mag} shows the dependence of the funnel wall configuration of 
 $ p = 1.0, \, 1.3, $ and $1.7$.  Hence,  one can understand the dependence of the magnetic 
 field shape on the acceleration efficiency of  jets. In Figure~\ref{fig:M87-mag}(a), 
 we plot the radial velocity $u^r$ for magnetic field lines within the conical boundary wall of 
 $\theta=\theta_0$.  Figures~\ref{fig:M87-mag}(b) and~\ref{fig:M87-mag}(c) show the cases 
 of parabolic boundary wall of $p=1.3$ and~$1.7$, respectively. When the jet's boundary 
 wall is conical ($p = 1.0$), it is estimated that $\Omega_F \approx 0.02/m$, whereas when 
 it is a parabolic shape, the value of $\Omega_F$ is overestimated compared with the case 
 of conical; i.e., $\Omega_F \approx 0.05/m$ for $p =1.3$, and $\Omega_F \approx 0.09/m$ 
 for $p =1.7$. 
 By fitting with KaVA data within several times $r_{\rm F}$, a value of about 
 $\Omega_F = (0.02 - 0.05) /m $ is obtained such that it should be slightly parabolic 
 for a slowly rotating BH magnetosphere according to Thoelecke et al. \cite{TST17}.
 Thus, the model of the magnetic line shape of $\Psi(r,\theta)$ and $\xi^2(r ;\Psi)$ used in 
 this study incorporates a plausible situation based on the observation results to some extent. 

%%\smallskip 

Although Hada et al.  \cite{Hada+13} observed bending of the jet shape, the physical reason 
for this bending was discussed in TT03. TT03 considered the flow of $E = \gamma_\infty \gg 1$, 
derived the approximated transfield equation for the outer region of the outer light surface,  
and solved the self-consistent magnetic field configuration. 
The outflow ejected from the plasma source was initially magnetically dominated. In the process 
of passing through the Alfv\'{e}n and fast-magnetosonic surfaces and accelerating,  the energy 
conversion from magnetic field energy to fluid kinetic energy occurs slowly but effectively. 
The bending of the magnetic field lines occurs in the area where the energy of both become  
equipartition. In other words, in the inner region where the magnetic field is dominant, it is 
conical-like or slightly parabolic, and when the plasma is accelerated and the plasma inertia 
becomes effective, the toroidal component $B_\phi$ of the magnetic field is enhanced. 
Therefore, the magnetic field lines are curved toward the rotation axis. Although the shape of 
the magnetic field lines in the vicinity of the BH is still unexplored, the observation results of 
the jet shape (i.e., magnetic field lines) of Hada et al.\cite{Hada+13} in the region close to the 
footpoint of the jet seem to roughly correspond to the magnetic field configuration suggested 
by TT03.

%%\smallskip 

 Notably, comparing the parabolic shape with the conical shape, for the same value of 
 $\Omega_F$, the location of the outer light surface along the jet, $Z_{\rm L}(\Psi_0)$, 
 shifts in the $+Z$-direction; however, the width of the outer light cylinder, $R_{\rm L}$, is 
 the same. Even if the $Z_{\rm L}(\Psi_0)$ is observably estimated along the streamline 
 of the jet, there is a dependence on both $\Omega_F$, $p$, and $\theta_0$. 
 If the velocity distribution in the width direction of the jet is measured, the width $R_{\rm L}$ 
 (i.e., $\Omega_F$) can be directly estimated. 

%%\smallskip 

\subsection{ M87 Jet Power and BZ Power }      \label{sec:BZpower}

 The BZ power is obtained from estimating the magnetic field strength near the rotating BH 
 and the angular velocity values (i.e., $0 < \Omega_F < \omega_{\rm H}$). It is difficult to 
 estimate the value of the BH spin from the observation data by Park et al. \cite{Park+19}. 
 However, the $\Omega_F^{\rm outflow}$ value can be estimated from the  observation and 
 by assuming the $\Omega_F$ values match in the outflow and inflow across the plasma 
 source region without significant energy loss. Hence, it can be considered whether BZ power 
 can be applied as the jet power. For example, we can assume the junction condition at the 
 plasma source as $\Omega_F^{\rm inflow} \approx \Omega_F^{\rm outflow}$, and  
 ${\cal E}^r_{\rm H} \approx  (\eta E)_{\rm outflow}$,  where  
 $ {\cal E}^r \equiv T^{r\alpha}_{\rm EM} k_\alpha$ is the radial component of the conserved 
 electromagnetic energy flux (i.e., BZ power) and $\eta E$ is the energy flux of MHD flow.  
 For the outflow where $\eta_{\rm outflow}>0$ and $E_{\rm outflow}>0$,  $(\eta E)_{\rm outflow}$ 
 indicates the jet power per magnetic flux.   
 Note that  the jet power $(\eta E)_{\rm outflow}$ comprises the magnetic and fluid parts of 
 the energy flux, whereas ${\cal E}^r_{\rm H}$ is purely magnetic energy flux from the BH.   

 %%\smallskip 
 
 We can also consider the case of 
 ${\cal E}^r_{\rm H} > (\eta E)_{\rm inflow} \approx (\eta E)_{\rm outflow}$ across the plasma 
 source, where $(\eta E)_{\rm inflow}$ can be obtained as a solution of type IIc negative 
 energy MHD inflow; i.e., $E_{\rm inflow} < 0$, $L_{\rm inflow} < 0$, $\eta_{\rm inflow} < 0$ 
 inflow \cite{TNTT90,Komisarov+04} (see also, \cite{GL14a} for gamma-ray bursts jets).   
 Notably, ${\cal E}^r_{\rm H} \approx (\eta E)_{\rm inflow} > 0$ in the magnetically dominated 
 limit (i.e., force-free case).  For the solution of type~IIc inflow, the magnetic field geometry 
 of Eq.~(\ref{eq:xi2-2}) should be adopted to avoid the break at the corotaion surface during 
 accretion onto the event horizon.  Moreover, the boundary conditions at the plasma source 
 that agree with the outflow solution must be estimated within the physical parameter ranges 
 such that $A(r,\theta) > 0$ between the plasma source and the event horizon.   
 
 %%\smallskip 
 
 When the magnetic field shape is conical near the event horizon, we have  
\begin{eqnarray}     
    L_{\rm BZ} &=& 2\pi  \int_0^{\theta_{\rm H}} \!\!\! d\theta ~
               {\cal E}^r_{\rm H} \Sigma_{\rm H} \sin\theta  \nonumber  \\ 
        &=& 2\pi  \varepsilon_0   
                \Psi_{\rm eq}^2  
                \int_0^{\theta_{\rm H}} \!\!\! d\theta ~ 
                \frac{ 2mr_H }{ \Sigma_{\rm H} } \Omega_F(\omega_{\rm H} - \Omega_F) 
                \sin^3\theta ~ ,  
               \nonumber  \\ ~                
\end{eqnarray}         
where $\theta_{\rm H}$ is the half opening angle of the jet at the event horizon. The magnetic 
flux function at the event horizon is assumed to 
$\Psi(r_{\rm H}, \theta) = \Psi_{\rm eq} \, ( 1 -\cos\theta )$ with a constant $\Psi_{\rm eq}$.   
The jet's opening angle is narrow enough, $\theta_{\rm H} \ll 1$, such that the BZ power is 
estimated as follows: 
 \begin{eqnarray} %{equation}     
     L_{\rm BZ} &\approx&  3.3 \times 10^{38} ~\Lambda ~ 
           \frac{ \Omega_F }{ \omega_{\rm H} } \left( 1 - \frac{ \Omega_F }{\omega_{\rm H}} \right) 
           \left( \frac{ a }{ m } \right)^2    \nonumber  \\ 
     & & \hspace{1cm} \times 
            \left( \frac{ B_{p{\rm H}} }{ 0.1 {\rm T} } \right)^2 
            \left( \frac{ m }{ 10^9 m_\odot } \right)^2 ~ \mbox{\rm  J s$^{-1}$ } ,  
 \end{eqnarray} %%{equation}     
 where $\Lambda$ is the geometrical  factor given by 
   \begin{equation}
       \Lambda = \left( \frac{2}{3} - \frac{3}{4}\cos\theta_{\rm H} + \frac{1}{12} \cos(3\theta_{\rm H})  \right)  
              ~ = \frac{ \theta_{\rm H}^4 }{ 4 } + O(\theta_{\rm H}^6) ~ . 
   \end{equation}
 Although we estimate $\theta_0 \sim 1/\hat{E} \approx 0.1$ rad at the distant region from the 
 BH,  in the vicinity of the event horizon, the magnetic field lines for inflows would be parabolic 
 \cite{TST17,TTS19} such that $\theta_{\rm H} > \theta_0 \sim 1/\hat{E}$.  So, when 
 $\theta_{\rm H} \approx 0.3$ rad at the event horizon,  we have $\Lambda \approx 0.002$ 
 as a rough value.  Hence, for $\Lambda = 0.002$, $B_{p{\rm H}} = 0.1$T, 
 $m=m_{\rm M87} = 6.5 \times 10^9 m_\odot$, $a = 0.9m$, and 
 $\Omega_F = 0.07 \omega_H = 0.0219/m$, we have 
 $L_{\rm BZ} \approx 1.5 \times 10^{36}$ J  s$^{-1}$,   
 which agree with the value estimated by the jet observation. 
 In Figure~\ref{fig:M87-KAVA}, the value of $\Omega_F$ was estimated for the conical 
 boundary wall; however, for the parabola-like boundary shape, the estimated $\Omega_F$ 
 value is about a factor larger. The BZ power may be an order of magnitude larger.  

%%\medskip 

\section{ Concluding Remarks } %================================

 In this article, we discussed the parameter dependence on the MHD outflows, and we realized 
 that the solutions of $u^r_{\rm inj}>0$ at $r_{\rm inj}=r_{\rm sp}$ or $r_{\rm sp} < r_{\rm inj}$ 
 with $u^r_{\rm inj}=0$. That is, the initial velocity of the flow solution is not always $0$, and the 
 plasma sources may be widely distributed between the inner and outer light surfaces.   
 Thus, the plasma source will have some internal structure; therefore, we should discuss 
 some physical process for the plasma source. It may be some kind of plasma instabilities, 
 nonideal MHD process, or particle pair-creation process   
\cite{Moscibrodzka11,Wong21,GL14a,GL14b,BIP92,HO98,HP16,LR11,BT15,Kisaka+20}.  
By connecting the outflow solution obtained from the observation data to the plasma source,  
we can consider the junction condition or restrictions there (i.e., \cite{Pu+15,PT20}). 
Further, we would discuss the inner BH magnetosphere. These details are outside the scope 
of this study and are for future study. 

%%\smallskip 

 In this study, we define the nondimentional magnetic field ${\cal B}_p$ and ${\cal B}_\phi$,  
 and introduce the ratio $\beta(r;\Psi)$ as a physical model. Therefore, there is no discussion 
 about the particle number flux number $\eta(\Psi)$ because this parameter is used 
 to make ${\cal B}_p$ and ${\cal B}_\phi$ dimensionless, which is related to the fact that 
 the critical condition at the magnetosonic points is no longer necessary in this model.  
 To estimate $\eta$ value, it is necessary to measure not only velocity distribution 
 $u^\alpha(r,\theta)$ but also magnetic field strength distribution $B_p(r,\theta)$ and 
 density distribution $n(r,\theta)$.  
 Therefore, it is paramount to understand the mechanism of shock wave formation 
 and the radiation mechanism in the jet region.  Thus, by comparing observational data, 
 it is possible to quantitatively discuss the particle number flux $\eta(\Psi)$. 

%%\smallskip  %%####################################################### 

 In summary,  
 we applied the trans-fast magnetosonic outflow solution in a BH magnetosphere to 
 the M87 jet. Specifically, we discussed the flow velocity $u^r(Z;\Psi)$ and the magnetization 
 parameter $\sigma(Z;\Psi)$.  We evaluate the values of field-aligned flow parameters;  
 $\hat{E} \approx 10$, $\tilde{L}\Omega_F \approx 0.9$, $\Omega_F \approx (0.02 - 0.05)/m $, 
 i.e., $\Omega_F \approx (20 - 50)$ year$^{-1}$. Using these values, the width of the outer 
 light surface (light cylinder) is roughly $R_{\rm L} \approx (20 - 50) m$, and the width of 
 the Alfv\'{e}n surface is roughly $R_{\rm A} \approx 0.95 R_{\rm L}$. As the spatial resolution 
 will be improved by future VLBI observations and the spatial distributions of jet velocity 
 $u^r(R,Z)$, and the magnetic field strength are known, the distribution and limits of 
 $E(\Psi)$, $L(\Psi)$,  $\Omega_F(\Psi)$, and $\eta(\Psi)$ will  be revealed. 
 Further, the future observations of the BH shadow by sub-millimeter wave and observations 
 via other wavelengths, such as X- and gamma-ray, will also impose restrictions on the 
 physical quantities of outflow and inflow. Thus, the quantitative understanding of BH 
 magnetosphere including the plasma sources; hence, the BH spacetime, will be deepened.

\acknowledgments

The authors thank Honoka Daikai for fruitful discussions. 
M.T. was supported in part by JSPS KAKENHI Grant No. 17K05439. 
M.K. was supported in part by JSPS KAKENHI Grant Numbers JP18H03721 and JP21H01137.
H.-Y. P. acknowledges the support of the Ministry of Education (MoE) Yushan Young Scholar 
Program, the Ministry of Science and Technology (MOST) under the grant 110-2112-M-003-007-MY2, 
and National Taiwan Normal University.

% For non-BibTex:
{}

\end{document}